\documentclass[journal,twocolumn,letterpaper]{IEEEJERM}

\usepackage{cite}

\ifCLASSINFOpdf

\else

\fi

\usepackage{setspace}
\usepackage{hhline}
\usepackage{multirow}
\usepackage{multicol}
\usepackage{array}
\usepackage{xspace}   % for smart macros' trailing spaces
\usepackage{graphicx} % for subfigures
\usepackage[tight]{subfigure}
\usepackage{paralist} % for concise lists
\usepackage{verbatim} % for comment blocks
\usepackage{url}      % for pretty URLs and better line-breaking
\usepackage{wrapfig}  % for within-text-flow wrap-around figures
\usepackage{booktabs}
\usepackage{colortbl}
\usepackage{tabularx}
\usepackage{amsmath}
\usepackage{color}
\usepackage{balance}
\usepackage{comment}
\usepackage[justification=centering,small]{caption}

\usepackage{setspace}
\usepackage{algorithm}
\usepackage{algpseudocode}
\usepackage{mathtools}

\renewcommand{\baselinestretch}{0.9}

\newcommand{\sysname}{ModelSink}

\begin{document}

\title{Communication and Synchronization of Distributed Medical Models: Design, Development, and Performance Analysis}

\author{Mohammad Hosseini,~\IEEEmembership{Senior Member,~IEEE,}
        Richard Berlin,~\IEEEmembership{Fellow,~IEEE,}
        Lui Sha,~\IEEEmembership{Fellow,~IEEE,}
        Axel Terfloth,~\IEEEmembership{Senior Member,~IEEE,}
        and~Houbing Song,~\IEEEmembership{Senior Member,~IEEE}% 
}
\markboth{IEEE Journal of Translational Engineering in Health and Medicine}
{M. Hosseini \MakeLowercase{\textit{et al.}}: IEEE Journal of Translational Engineering in Health and Medicine}

\twocolumn[
\begin{@twocolumnfalse}
  
\maketitle

\begin{abstract}
Model-based development is a widely-used method to describe complex systems that enables the rapid prototyping. Advances in the science of distributed systems has led to the development of large scale statechart models which are distributed among multiple locations. Taking medicine for example, models of best-practice guidelines during rural ambulance transport are distributed across hospital settings from a rural hospital, to an ambulance, to a central tertiary hospital. Unfortunately, these medical models require continuous and real-time communication across individual medical models in physically distributed treatment locations which provides vital assistance to the clinicians and physicians. This makes it necessary to offer methods for model-driven communication and synchronization in a distributed environment.

In this paper, we describe ModelSink, a middleware to address the problem of communication and synchronization of heterogeneous distributed models. Being motivated by the synchronization requirements during emergency ambulance transport, we use medical best-practice models as a case study to illustrate the notion of distributed models. Through ModelSink, we achieve an efficient communication architecture, open-loop-safe protocol, and queuing and mapping mechanisms compliant with the semantics of statechart-based model-driven development. %, and employ a synchronized atomic queuing model to ensure the real-time requirements for synchronization are met. We also develop parameter mapping mechanisms to automatically share synchronization control events from one model to another, and implement an open-loop safety mechanism ensuring that interactions among multiple distributed medical models function in a safe and reliable manner. 
We evaluated the performance of ModelSink on distributed sets of medical models that we have developed to assess how ModelSink performs in various loads. Our work is intended to assist clinicians, Emergency Medical Technicians (EMT), and medical staff to prevent unintended deviations from medical best practices, and overcome connectivity and coordination challenges that exist in a distributed hospital network. Our evaluations are based on real-world case studies done by research community, model-driven software industry, and hospital as the end user. Our practice suggests that there are in fact additional potential domains beyond medicine where our middleware can provide needed utility, such as automobile and avionics industry, e-learning, or any other domains where statechart models are employed.
\end{abstract}


\end{@twocolumnfalse}]

% Put footnotes here
{
  \renewcommand{\thefootnote}{}%
  \footnotetext[1]{Department of Computer Science, University of Illinois at Urbana-Champaign}
  \footnotetext[2]{Department of Surgery, Carle Foundation Hospital}
  \footnotetext[3]{Yakindu Statechart Tools, Itemis AG}
  \footnotetext[3]{Department of Electrical, Computer, Software, and Systems Engineering, Embry-Riddle Aeronautical University}
}
 
\IEEEpeerreviewmaketitle

\section{Introduction and Background}
\label{section:introduction}
Medical best-practice guidelines, instructions and standards of care play a vital role in medical care. Results of studies report that 44,000 to 98,000 Americans die each year as a result of failing to follow medical best-practice guidelines \cite{guideline1, guideline2}. Many medical best-practice guidelines exist in hospital handbooks that are commonly lengthy and quite difficult to apply clinically, particularly in the acute medical care setting. Therefore, more and more text-based medical best-practice guidelines are represented and encoded in computer-interpretable formats such as Arden \cite{arden}, GLIF \cite{glif}, and PROforma \cite{proforma}. Decision support systems such as Spock \cite{spock} have been developed to monitor treatment decisions and provide medical staff with proper, timely recommendations. However, most such encodings and representations are similar to the format of executable pseudo code, which are usually at a quite low level for medical staff. Many clinical problems are complicated and such formats do not provide a visual and user friendly interface for physicians to validate correctness. Furthermore, it is not easy to verify formally these formats using the requirement of rigorous correctness needed for life-critical medical cyber-physical systems.

Advances in techniques of software engineering accelerated the industrial use of model-driven development, and visual modeling in particular, for building very large and complex models not only in the medical domain, but also other domains such as automobile and avionics \cite{complexSoftware, visual, statecharts2}. It is said that 60-90\% of production in the automotive domain for example, is done through model-driven development \cite{complexSoftware}. Organizations such as FDA have supported initiatives to aid in the effective development medical devices %The Center for Devices and Radiological Health (CDRH) branch of FDA for example, supports contributions aimed for effective development techniques of medical devices. Multiple public workshops focusing on promoting advances in medical device modeling and simulation have been organized by FDA for this purpose 
\cite{fda1, fda2}. For medicine, the fact is that visual models such as statecharts are very similar to disease models and treatment models, and are executable and can be indirectly verified motivates the use of statecharts for encoding medical guidelines. Their well-designed user interface, simulation functionality, and hence rapid prototyping and validation helps medical staff to understand the design more easily, validate the design model through user-friendly simulation, and therefore give more meaningful suggestions to the model of the best-practice guidelines.

\begin{figure}[!tbp]
    \centering
        \includegraphics[width=.8\columnwidth]{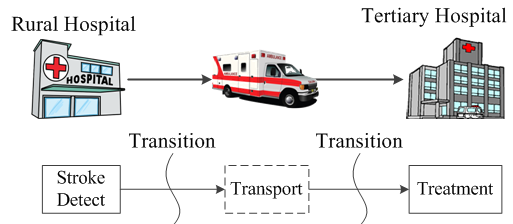}
    \caption{Current stroke management inherently distributes care provision from remote rural to ambulance to central hospital.}
    \label{Fig:Distribution}
    \vspace{-.5cm}
\end{figure}

Emergency patient transport for acute diseases from a rural hospital to a tertiary center hospital such as those seen in stroke scenario has naturally complicated the process of adherence to best practice guidance. Executable medical best-practice models are distributed across multiple locations from the time that a patient first presents at a rural hospital, through diagnosis and ambulance transfer, to arrival and treatment at a regional tertiary hospital center. This has led to the creation of medical models that are widely separated physically and must run synchronously while communicating with each other in real-time and sharing clinical data as necessary. These distributed medical models are executable statecharts based on various disease models, patient condition models, and models of care provision capability. Care provisions vary from remote rural to ambulance to central tertiary referral. Figure \ref{Fig:Distribution} shows an abstract view of a stroke rural patient transport. Best-practice models are executed in real-time at each of the three locations, with physicians at the tertiary center facility supervising the remote rural hospital physicians and/or the EMT in the transporting ambulance. This approach extends in a virtual manner the boundaries of the tertiary center to include the remote facility and transport channels using models and an integrated communication system that safely and effectively synchronizes care across large hospital network. The models of patient and system remain consistent throughout. Furthermore, the heterogeneous nature of medical models exacerbates the problems as model design applications may lack consistency and common features of design. This makes the problem of monitoring, traceability, and validation even harder for the medical software developers during development of medical systems \cite{joms, hosseinidataset, chase17, hosseiniRouteScheduler}.

%Without loss of generality, similar problem is extended to the context of interoperability and coordination of various medical devices. The lack of coordination and synchronization among medical devices is a daily problem across hospital networks \cite{devices1, devices2}. Medical devices such as ventilators, IV pumps, heart monitors and computers holding patient records communicate and update one another and get synchronized automatically \cite{devices1, devices2}. Unfortunately a big problem in hospitals is that most of these medical devices can't communicate with eachother to share data. A recent survey on 526 nurses \cite{survey} showed that more than half of the nurses had witnessed a medical error caused by a lack of coordination among devices. More than two-thirds of the nurses said they spend at least an hour a shift dealing with communication issues among devices, troubleshooting problems, and recording data manually that isn't shared among devices automatically. Such problems place a heavy burden on nurses to translate and exacts a toll on patients as well. 93 percent of nurses in the survey strongly agreed that medical devices such as monitors, diagnostic devices, and others should be able to seamlessly share data with one another automatically \cite{survey}.

The problem of coordination and synchronization may sometimes be hazardous that needs to be considered during the design and modeling phases. In laser tracheotomy, for example \cite{laser1, laser2}, a surgeon operates a laser scalpel to unblock the airway of the patient. In a simplified scenario, the system has two main components: the ventilator to supply the oxygen or plaint air, and the laser scalpel to emit the laser. Coordination among these two components is highly critical to ensure the ventilator and the laser scalpel are not operating at the same time. More specifically, laser scalpel should not start operation until a certain amount of time past from shutting off the ventilator to make sure the amount of oxygen left in the patient is always kept below a specific safe threshold. Similarly, to avoid brain damage due to hypoxia\footnote{Hypoxia is a condition where the oxygen concentrations fall below the level necessary. Hypoxia may cause severe brain damage or brain death.}, the ventilator should not be kept at non-operational status for a specific amount of time. Overall, a need for an intermediary system to help medical devices software or their executable models to communicate with each other, and to update, coordinate, and to get consistent with each other is therefore evident.

%This makes the interaction and concurrent subjective functions such as traceability, validation, and best practice recommendation difficult for the medical staff co-monitoring the synchronous statechart models. Besides, the heterogeneous nature of the statechart models exacerbates the problem as the third-party applications used to design the models and their design purposes can be different. For instance, one model can be a verified timed automata-based model (e.g. in UPPAAL) while others may be stateflow models (e.g. in Simulink) designed for simulation purposes. Both of these two reasons motivate the need for co-simulation among different distributed models \cite{distributed2}.
%benefits: http://www.theenterprisearchitect.eu/blog/2009/11/25/15-reasons-why-you-should-start-using-model-driven-development/
\begin{figure*}[!tbp]
\vspace{-.3cm}
    \centering
        \includegraphics[trim=6.7in 3.2in 6.52in 3in, width=1.01\textwidth, height=0.36\textwidth]{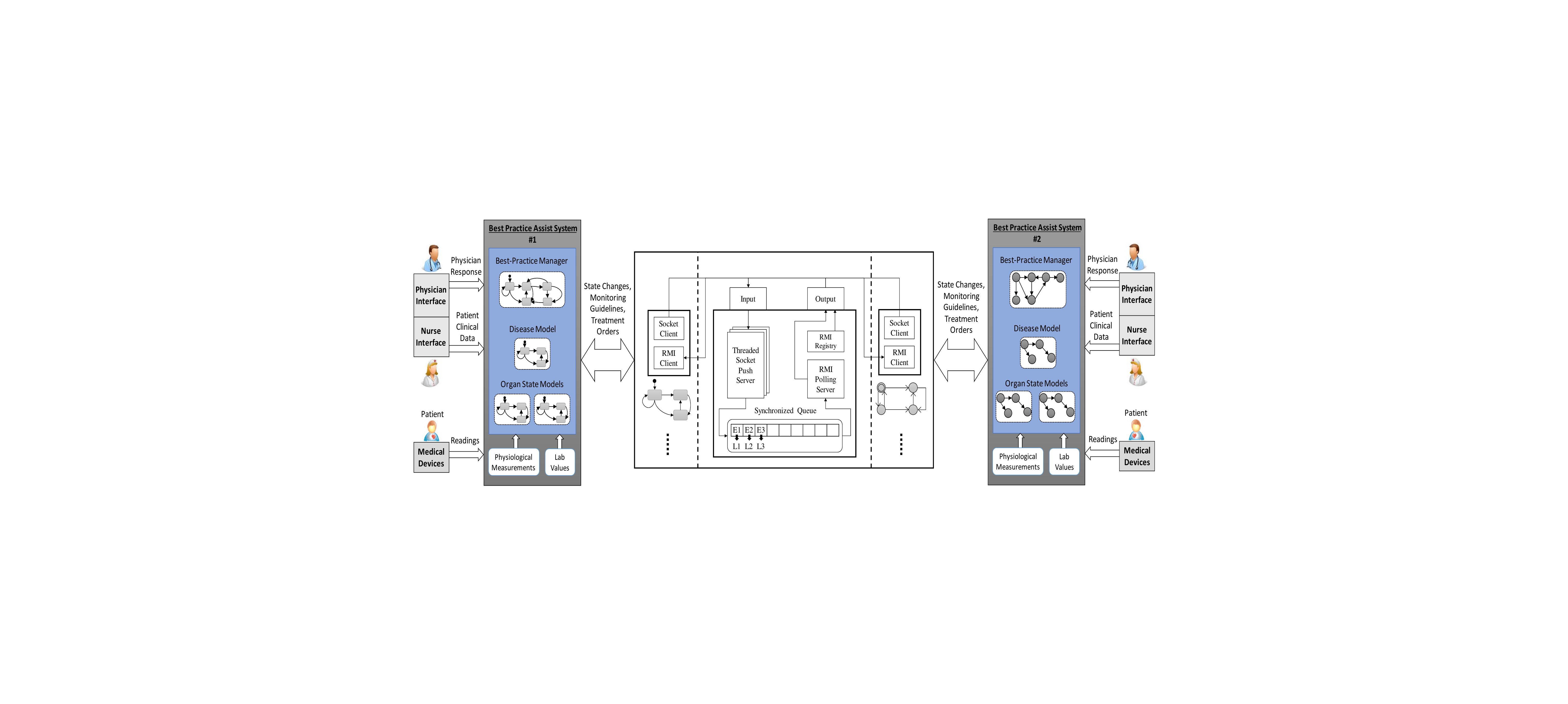}
    \caption{The \sysname\ workflow. The middleware allows heterogeneous distributed executable medical models (medical best-practice models in this example) to communicate and synchronize automatically.}
    \label{Fig:Workflow}
    \vspace{-.5cm}
\end{figure*}

In our preliminary work \cite{joms, iccpswip}, we ran a pilot study on a pathophysiological model-driven communication for dynamic distributed medical best practice guidance systems. 
%During this study, we proposed the notion of distributed medical best-practice guidance systems across multiple distributed hospital networks, the dynamism introduced by this distribution, as well as the concept of executable best-practice automata. We targeted acute stroke patient emergency care as a use case, and codified medical best-practice knowledge available in medical textbooks into executable automata. We presented an initial design of a message-exchange architecture, and introduced our customized pathophysiological model-driven communication protocol.
 
The work presented here follows our pilot study, and builds upon our earlier model-driven design towards a real-platform development and implementation of a model-driven communication middleware for synchronization of the executable models that we codified during our pilot study, followed by performance evaluation through extensive experiments and instrumentation. We describe \sysname , a middleware that enables automatic model-to-model communication and synchronization among heterogeneous statechart models, not only during execution, but can also aid model-driven development, simulation, validation, and traceability of any distributed statechart models. Through ModelSink, distributed statechart models can communicate and be synchronized with each other efficiently through automatic sharing of data such as patient state changes for example in the clinical context. This permits the communication and automatic synchronization of various statechart models such as medical best-practice guidance models in distributed hospital settings, in an ensemble. In summary, ModelSink 

\begin{enumerate}[i]
\item automatically performs model-to-model synchronization by sharing state changes as opposed to manual updates by the users. ModelSink therefore facilitates communication among physically distributed users (e.g. EMT and doctors at a center hospital) to share status updates.
\item leverages an efficient communication architecture to distribute messages among disjoint statechart models. Our design uses a customized low-overhead persistent push/poll communication mechanism which also complies with the semantics of automata-based model-driven development.
\item leverages a synchronized atomic and thread-safe queuing module to achieve the real-time requirements for co-execution and synchronization of statechart models. Our specialized design removes possible synchronization overhead between the pushes and the multi-threaded polls.
\item provides a distributed mechanism to avoid life-critical safety issues within the context of distributed medical systems when communication fails. We embed an open-loop safety parameter field in the communication protocol header to ensure that the models \textit{always} transit to a safe state in case of communication failure or message loss.
\item leverages a parameter mapping module to transfer messages among distributed models for the purpose of supporting coordination and synchronization. The module helps translate model-specific synchronization control events from one model to another.
\item is platform independent. ModelSink can be deployed on any platform, and supports the heterogeneous nature of statechart models as the design semantics are independent of the underlying modeling application.
\end{enumerate}

To the best of our knowledge, this is the first middleware that achieves automatic communication and synchronization among heterogeneous distributed statechart models.

Figure \ref{Fig:Workflow} illustrates a view of the \sysname\ workflow. The heterogeneous medical models such as best-practice guidance models monitored by physicians and nurses are executed in different locations. Once the medical models are connected, our middleware automatically shares synchronization data and control events such as patient state changes and monitoring guidelines in the context of best practices across the distributed medical and clinical models, remotely, as if they are manually controlled by the medical staff. Furthermore, \sysname 's automated communication and synchronization mechanisms achieved through \textit{model-only} control incurs a high degree of flexibility, and can effectively adapt to clinical ecosystem changes when reconfiguration of medical models such as disease models rapidly occurs. That leads to significant reduction in heterogeneous model-driven development and software maintenance costs of medical best-practice models.
%http://docs.spring.io/autorepo/docs/spring-statemachine/1.0.x/reference/html/appendices-zookeeper.html
\sysname ~can further assist software engineers to build a single user-friendly gateway with minimum complexity that can be used as a front-end interaction to realize co-simulation, co-execution, or to monitor and control large-scale statechart models simultaneously. 

%It should be noted that this paper is not intended to propose new treatments or discover new medical knowledge, but to assist clinicians, EMT, and medical staff to address difficulties in implementing best practices, and prevent unintended deviations and faster real-time adherence to medical best practices. Further, it also helps them overcome connectivity, consistency, and coordination challenges that exist in a distributed hospital network, such as those seen in patient transfer from rural hospital, to ambulances, to center hospital.

\section{Related Work}
\label{RelatedWork}
\sysname\ is conceptually similar to the notion of \textit{mediators} underlying emergent connectors \cite{mediator1,mediator2,mediator3} such as Enterprise Service Bus \cite{esb} as the concept of a ``connectivity middleware" is common between the two. However, \sysname\ is fundamentally different as the design goal of mediators is to enable the composition of pervasive networked systems and protocol mediation as opposed to remote synchronization of executable statechart models. Another set of related work to \sysname\ is database data replication tools such as \cite{db1,db2,db3} and \cite{db4} that provide automated data sharing and data replication between databases. Overall, while there are multitude of related work from computer science perspective, these tools have not been properly integrated into medicine and the safety-critical aspects of clinical practices, such as handling connection failure and safety in cases of communication loss. A major distinguishing point for \sysname\ therefore, is to adapting to the context, such as clinical needs as well as the life critical requirements of medical domain. Furthermore, in \sysname\ , on the contrary, the notion of access is platform independent and lightweight because it provides a \textit{model-to-model} access as opposed to \textit{system-to-system} or \textit{database-to-database}.

In the medical context, medical best-practice guidelines for emergency care have been created for patients in major hospitals. For instance, the University of Texas’ MD Anderson Cancer Center has developed clinical management algorithms \cite{anderson} that depict best practices for diagnosis, evaluation, and treatment of specific diseases such as cancer targeting adult patients. Their contribution however, only provides a high level algorithmic workflow using a multi-disciplinary approach, and not only are not modeled as executable guidelines, but also are conceptually centralized, and therefore the notion of communication, synchronization, and distribution is not applicable in their context.

A variety of heterogeneous model construction and interpretation have been proposed for the simulation of complex systems. Ptolemy \cite{buck1994ptolemy} for example, the most famous heterogeneous modeling platform, supports prototyping of heterogeneous systems with different domain computation model to capture different type of subsystems. Current examples of domains include synchronous and dynamic dataflow, discrete-event, and others appropriate for control software. Domains can be mixed as appropriate and use underlying meta model to realize an overall system simulation. Similar methods are adopted by \cite{jiang2015design} and \cite{radojevic2011design}, as they use automata to interpret heterogeneous timed automata and dataflow model and capture the heterogeneous sub-models. These works however, function in a single project and a single machine, and lack support of communication and distributed execution.

Some works have been done to address the problem of co-simulation for heterogeneous models located in different projects. For example, in \cite{fitzgerald2014co}, the authors implement Crescendo, a tool that allows the model expressed in discrete-event of a tool called \textit{VDM} and the model expressed in continuous time of a tool called \textit{20-sim} to share information and accomplish co-simulation among them. In \cite{blochwitz2011functional}, the authors propose FMI, a tool to support model exchange using a combination of XML files and C-code. Similarly in \cite{bombino2010heterogeneous}, authors propose a code-in-the-loop co-simulation framework for the OMG SysML implemented in the Artisan Studio tool, and the Stateflow model implemented in Matlab Simulink tool. The main feature of their tool is the automatic generation of optimized code, allowing simulation that may eventually run natively on a target for embedded systems. While these works capture the heterogeneous models, they are still limited to a single machine and lack support for distributed models, which need significant effort to solve the communication, synchronization, and consistency which motivates our work.

\section{Design of the Middleware}
The core of \sysname\ is implemented through a queuing and mapping system as well as a push-based and poll-based communication system, which are accomplished through socket-based and Remote Method Invocation (RMI)-based client/server architecture, respectively. Two client agents (a push socket client agent and a polling RMI client agent) are installed on every machine executing any statechart models. These clients are then connected to the corresponding servers that can be located on any of those machines. During \sysname\ sessions, all corresponding communication and synchronization control events are registered and distributed through \sysname\ as if medical staff were actually monitoring and manually tracking the disease or patient status through their executable models. We implemented \sysname\ in Java as a partial compliance with the design feature (vi), so that it can be deployed on any platform running Java Virtual Machine (JVM), including Linux and Windows, therefore, making the compiled code platform-independent. We have designed an open-loop safe communication protocol for cases where communication fails, as well as a list of APIs for ease of operations such as establishment of a connection, specifying medical control events and rules for synchronization, as well as specifying push or poll mechanisms for communication or synchronization control events.

\begin{figure*}[!tp]
\centering
\includegraphics[width=.8\textwidth]{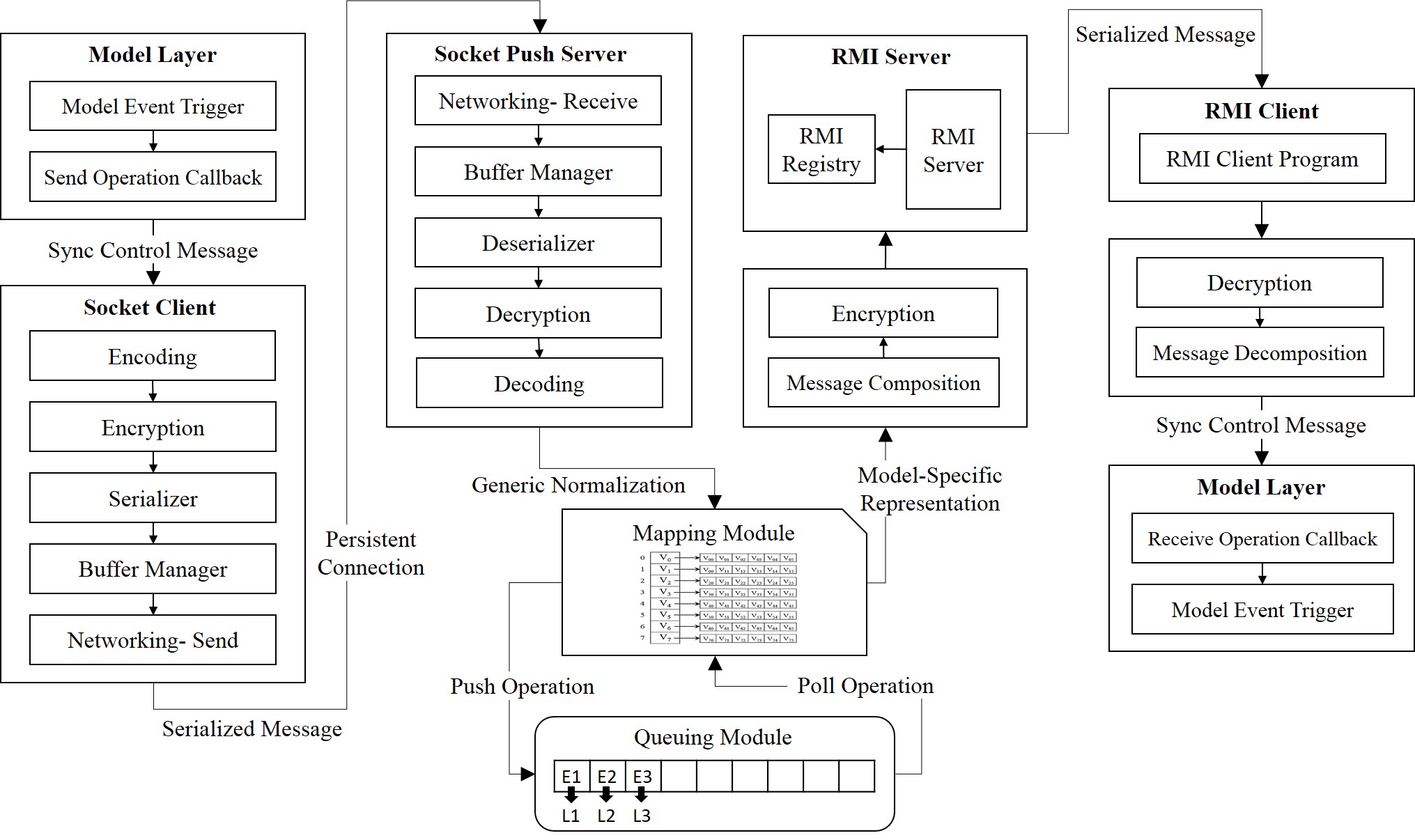}
\caption{The overall structure of \sysname\ .}
\label{Fig:Framework}
\vspace{-.5cm}
\end{figure*}

\subsection{Middleware Structure}
\sysname\ consists of six major components: a socket-based client agent (push client) and a RMI-based client agent (polling client), both residing on each medical machine, a mapping module for synchronization purpose, a queuing module, a multi-threaded socket-based server (push server) and a RMI-based registry and server residing on any of the distributed medical machines designated as the server. The overall structure of \sysname\ is illustrated in Figure \ref{Fig:Framework}.

The push client captures medical events through operation callback on the executable models once an event such as a state change is triggered. These locally-triggered synchronization control events are then encoded to a specific message format identifying the statechart model's UID and triggered events which overall forms the synchronization control message. The messages are then encrypted with the AES 128-bit symmetric cipher in Electronic Codebook (ECB) mode, serialized, buffered, and then sent to the push server via the persistent socket connection that has been established. The push server is a socket-based multi-threaded server that communicates with all other remote push-client agents concurrently. The server receives synchronization control messages associated with each executable statechart model, which are eventually pushed into the queuing module. Once received at the push server, the synchronization messages are deserialized, decrypted, decoded, and normalized to a generic format which are then directed to the mapping module. The use of the generic format is to store messages as a globally understandable format, which later helps with model-specific translation of synchronization control events specific to each statechart model.

To enable synchronization among distributed statechart models, the mapping module is pre-configured with semantic mappings of synchronization control events from one model to another, to provide a particular set of events specific to each of the distributed statechart models. This happens by performing translation of synchronization control events from one statechart model to corresponding control events of other statechart models located on other remote devices or platforms, thereby automatically synchronizing statechart models or co-executing medical devices altogether. While placement of the mapping module in a centralized manner on the server machine is more convenient for applying or spreading updates as well as for auditing and security purposes, it is not yet a hard requirement. The mapping module can be partially placed locally on each of medical machines alternatively.

The normalized synchronization messages originated from the mapping module are then pushed to the atomic thread-safe FIFO queuing module. Upon polling operations which are done through RMI-based communication system, the generic formatted messages are then dequeued from the queuing module, and are re-mapped to form model-specific events, and to generate a set of synchronization control events specific to each statechart model. Messages are then composed and deployed on our open-loop-safe communication protocol supporting both \textit{reliability} and \textit{safety} features, and are then sent to the RMI polling server. The synchronization control messages are deserialized and decrypted once polled by the RMI client agents residing on each statechart model. Once the synchronization control events are decomposed, synchronization among distributed statechart models is performed by triggering the destination model's \texttt{receive()} operation callback formed in accordance with the originating model's synchronization control events received through the communication channel.

\subsection{Communication Architecture}
In compliance with the design feature (ii) (refer to Section \ref{section:introduction}), we employ an efficient communication architecture inside ModelSink to propagate messages among distributed statechart models. Our design uses customized low-overhead persistent push-based and poll-based communication mechanisms which is also compliant with the semantics of automata-based model-driven development. From an engineering point of view, unlike a regular client-server communication such as those used in messaging applications with the client process looping around a buffer to read responses, our medical middleware must also support continuous and sporadic message transfer, but with no termination of the socket connection. However, it also needs to maintain safety, security, reliability, as well as a long-lived connection after each data transfer in order to incur minimum latency.

To address the communication requirements, we customized a low-overhead persistent socket-based client-server communication architecture over TCP/IP for push operations throughout the running sessions rather than setting up a new connection for each push operation. This maintains the stability of the socket connections by initially creating a connection at the beginning of each transfer session, and occasionally sending a message as necessary. To enable that, we wrapped the push client socket connection around a thread, and use a blocking queue to wait for messages. A single sender queue exists throughout the application, therefore using a singleton pattern. On the other hand, performing a $\texttt{read()}$ function causes the thread to block forever. To address that, we use a special thread that calls a specific method repetitively at specified periods and read time-out that can be used to post a ping message, once in a while. This improves the stability of connections while also relaxing problems associated with application terminations due to calling the $\texttt{close()}$ function.

\begin{figure}[!t]
\centering
\subfigure[Rural Hospitals]{
    \centering
    \includegraphics[width=1\columnwidth]{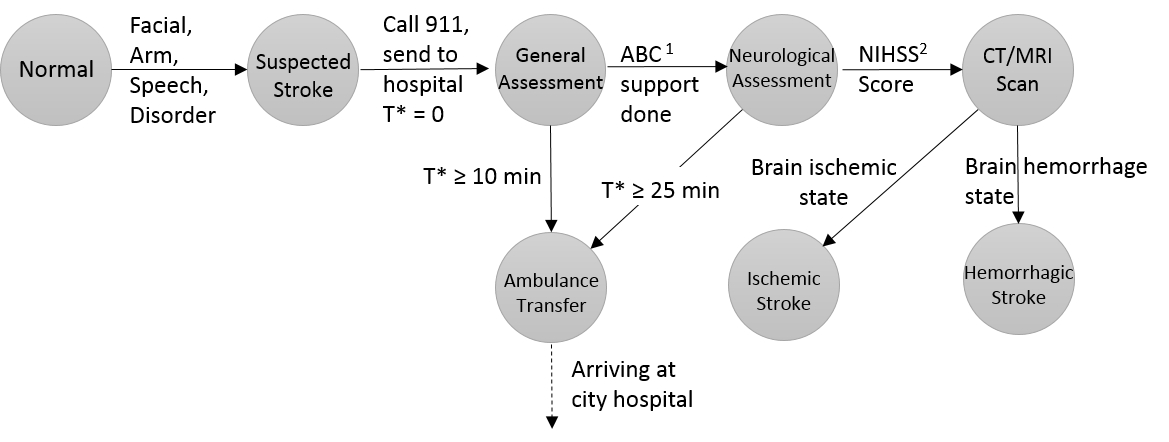}
	\label{fig:StrokeAutomataRural}}
\subfigure[Center Hospitals]{
	\centering 
	\includegraphics[width=1\columnwidth]{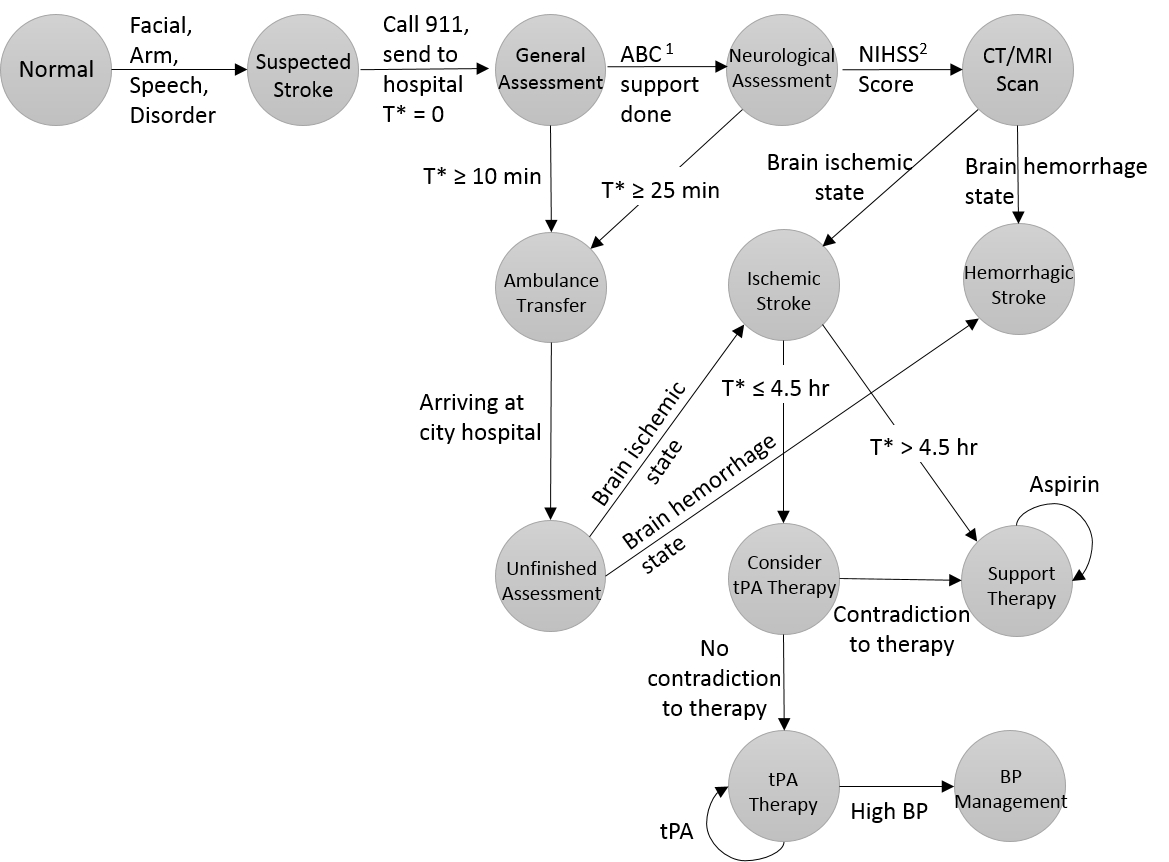} 
	\label{fig:StrokeAutomataCenter}}
\caption{A simplified version of best-practice workflow models for stroke, distributed between rural hospital (top) and regional tertiary hospital (bottom).}
\label{Fig:stroke}
\vspace{-.5cm}
\end{figure}

While the socket-based client-server architecture is good for individual statechart models to push messages to the socket-based server, this is not yet a good solution for distributing the messages between the distributed statechart models as it imposes a significant commitment on the server side. The server is forced to keep a long-lived and mostly unused socket connection open with each client in an inefficient way. This leads us to use client-side polling, for two main reasons: First, client-side polling is architecturally simpler. Using this approach, the server doesn't have to track which statechart clients called and which statechart clients are waiting for messages. This leads to simpler implementations while also making it easier to support various types of statechart clients; meeting the heterogeneous requirements of the distributed medical devices and their executable models. The second reason to use client-side polling is the inability of clients to accept socket connection initiated from server, especially in the context of automata-based model designs. Our design enables ModelSink to support the heterogeneous nature of statechart models and makes it independent of the automata modeling application, partially addressing the design feature (vi). For example, in models designed in Simulink's Stateflow, the destination model reads the value of data from an input port which is only possible through polling mechanism. Similarly, in models designed in Yakindu's Statecharts modeling software, it is not possible to directly raise events from the operation callback and trigger a receiving data. Nevertheless, the only option here is to poll values in a guard expression. Our use of RMI-based callbacks further helps with significant efficiency improvements such as decreased client-side processing as well as implementation simplicity and support of various statechart modeling frameworks \cite{grosso}. %Aside from these, if the client statecharts is an applet or when firewall is inlvolved, creating sockets is forbidden or the network traffic to certain sockets get restricted.
%https://books.google.com/books?id=MgS4bJKojaQC&pg=PT494&lpg=PT494&dq=why+RMI+is+good+for+polling

\subsection{Open-Loop Safety Protocol}
Communication failure in the wireless medical environment can lead to life-critical safety issues within the context of distributed medical systems. Following the design feature (iv), the architecture of our medical middleware should guarantee the safety of the execution of distributed medical statechart models, to ensure that the models \textit{always} transit to a safe state from a model-driven perspective even with communication failure or loss of messages. Let's take Stroke best-practice guidance models as an example. Figure \ref{Fig:stroke} illustrates a simplified version of our developed stroke best-practice workflow model both for rural hospital (top) and regional center hospital (bottom), which form the executable core of best-practice guidance models. Assume a message triggers a state change event in the model of regional center hospital, making the workflow model transit to ``tPA Therapy'' state. Suddenly communication failure occurs. A question which arises is ``how long to continue tPA therapy and stay in that state?''. Continuing tPA therapy for longer than a specific duration characteristically is hazardous for the patient, which therefore is considered to be unsafe for the whole cyber-physical-medical system. 

Similarly, let's consider the laser tracheotomy example again. As mentioned earlier, to avoid surgical fire, one of the vital safety requirements is to ensure the ventilator and the laser scalpel are not operating at the same time, and that laser scalpel does not start operation until a certain amount of time has been past from shutting off the ventilator. Similarly, to avoid brain damage due to hypoxia, the ventilator should not kept non-operational for a specific amount of time. In such scenarios therefore, continuing emission of laser or keeping ventilator off more than a specific period are both considered unsafe for the whole system.

Given the aforementioned characteristics, we therefore classify states of model execution into the following two classes:
\begin{itemize}
\item A transient safe state, which allows models to stay safely in the state, but only for a limited duration. That said, if staying on a transient safe state lasts longer than the specified allowed limit, it becomes unsafe, and may lead to medical hazards. ``tPA Therapy'' state is an example of a transient safe state.
\item An open-loop safe state, which is considered always-safe for the maximum duration of the given medical procedure. Therefore, an open-loop safe state does not involve any medical hazard while stay lasts more than any time threshold. 
\end{itemize}

\begin{figure*}[!t]
\centering
\includegraphics[width=1.05\textwidth]{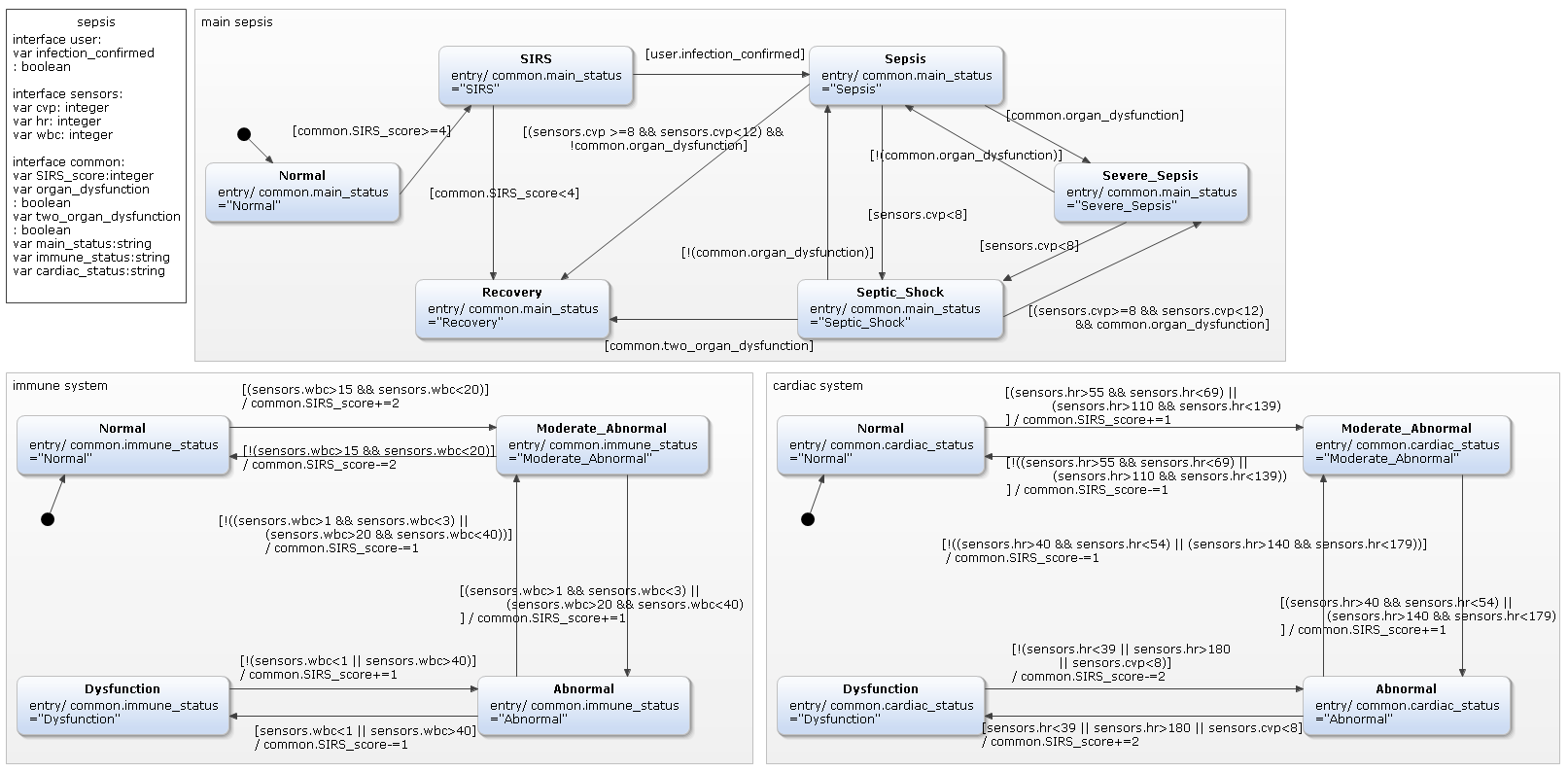}
\caption{Simplified statechart-based models for sepsis, including disease (top) and two of the underlying organ models (bottom).}
\vspace{-.5cm}
\label{Fig:sepsis}
\end{figure*}

To maintain reliability and safety, our designed communication and synchronization system uses open-loop safety to guarantee that execution of medical models transits from a transient-safe state to a predefined open-loop-safe state when a communication failure occurs. Earlier works on open-loop safety protocol in the context of a medical systems include \cite{openloopsafe1} and \cite{openloopsafe2}. The authors in these works used an airway laser surgery system as the target example to propose a \textit{centralized} supervisor which decides a safe operation region for medical devices. While these works share the same assumption that the communication network is unreliable, they deal with the problem of open-loop safety by a centralized approach. Whereas in our work, we employ the notion of open-loop safety mechanisms in a distributed medical environment, and extend that towards medical best-practice models in a distributed hospital network. As a part of tackling the problem in the distributed environment, we embed open-loop safety as a safety parameter field into our communication protocol's header, so that a communicated message triggering a state change forces the medical statechart models not to make a state change unless an open-loop state is already determined and queued as an emergency option in case communication fails. It is also necessary to assure that the open-loop safety option is executable locally on patient's side. In the context of Stroke best-practice guidance models, transient safe states such as ``tPA Therapy'' are transited to an \textit{implicit} ``general assessment'' as an open-loop-safe state to ensure the safety requirements of execution of distributed medical statechart models.

\subsection{Data Structures and Rules for Mapping}
Following the design feature (v), ModelSink leverages a parameter mapping module to transfer messages and control events among distributed executable models for the purpose of supporting coordination and synchronization. It provides a mapping of model-specific synchronization control events specific to each individual model from one model to another. The mapping module works on the principle of key-value store and hashing, composed of a combination of multi-hashmap and 2-dimensional linked-list data structures, which is pre-configured with mappings of synchronization control events from one statechart model to another for the purpose of coordination and synchronization. To store key-value pairs, we used the first dimension of the 2D linked-list as a bucket to store key objects corresponding to encoded normalized generic synchronization control events. The second dimension is used to store values corresponding to an ordered list of model-specific synchronization control events specific to each individual model such as necessary medical functions and state changes that must be triggered on the other statechart models. Similar to a regular \textit{HashMap}, the mapping module's $\texttt{get(Key k)}$ method calls \textit{hashCode} method on the key inputs, and applies returned \textit{hashValue} to its own static hash function to find a bucket location where keys and values are stored.

Our implementation of the mapping module imposes a one-time overhead, while it can also be reusable. Therefore, if the specifications of statechart models change considerably, only the mapping module is updated, therefore incurring minimum cost.

\subsection{Data Structures and Rules for Queuing}
According to the design feature (i) and (iii), to achieve the real-time requirements of co-execution and synchronization of emergency medical systems, it is crucial to remove possible synchronization overhead between the multi-threaded push server and the pipelined RMI polls. As mentioned earlier, for higher performance, \sysname\ does not rely on send/receive messaging model to distribute messages across different statechart clients. Instead, statechart clients retrieve messages by \textit{polling} data from a shared synchronized queue-like data structure where only the socket push server writes. This specific approach is called \textit{pointer polling} and is more efficient, is simpler, and requires no additional storage data structure on the receiving side compared to \textit{send} approach when data is continuously produced.

In order to synchronize multiple pipelined RMI polling accesses, we developed a \textit{listed chasing-pointer queue} which is our efficient customized list. Our \textit{listed chasing-pointer queue} is a specialized data structure which employs an efficient wait-free algorithm to support low-latency concurrent accesses. We borrow concepts from \cite{queue1} and \cite{queue2} to implement a variation of an efficient algorithm that can be employed for \textit{multiple} number of producers and consumers. The queue is thread-safe, and is based on linked nodes. The queue orders messages FIFO, with the head of the queue being the message pushed the longest time and the retrieval operations obtaining elements at the head of the queue. The traditional implementation of a synchronized queue works based on the concept of a shared variable such as a \textit{counter} to synchronize the writer and the reader, which is referred to as a counter-based approach. Since both the writer and the different readers modify this shared variable, multiple cache misses are unavoidable for distribution of each message on the server side, therefore resulting in significant performance overhead when synchronization has to be performed very frequently. Our specialized implementation avoids use of a shared synchronization variable, and instead uses a list of boolean flags to indicate whether a message is available in the list to poll which reduces the synchronization issues. Each message slot in the queue is augmented with a header that indicates whether a message is available or not. Each queue cell contains the normalized generic message value, which points to a list of flagged pointers, with each index associated with a specific statechart client. Thus, a consumer statechart client is always chasing the producer statechart client in the synchronized queue for filled queue cells. With that feature being implemented, every statechart client regularly polls only its own pointer, and unflags it whenever data is successfully retrieved. The process of polling messages from the queue continues until at least one pointer associated with a specific statechart model is still flagged. Once all pointers are unflagged, the data is dequeued, and the head of the queue is advanced.
%https://docs.oracle.com/javase/7/docs/api/java/util/concurrent/ConcurrentLinkedQueue.html#offer(E)
%\subsubsection{s} OPEN-LOOP SAFETY, APPLICATION PROTOCOL HEADER

\section{Evaluation}
From a medical perspective, physicians are taught organ system function as part of the representation of disease process. They look for patterns of pathophysiological changes (the change in physiological measurements as a result of disease) within an organ system to understand organ states \cite{organ1, organ2}. This organ-centric view of pathophysiological expression also matches medical treatment, which is captured by a best-practice medical system. Given that, to evaluate our middleware, we developed multiple medical best-practice executable models, and used them as benchmarks and proof-of-concepts to evaluate ModelSink. The engine of our best-practice guidance systems is executable medical statechart models including disease and organ models. We codified medical knowledge into executable formal best-practice models so that they can be checked by expert physicians via the execution of these models using scenario-driven simulation and a user-friendly graphical interface that the statechart-based design provides.

\subsection{Experimental Setup}
For the development of the executable models, we used Yakindu Statechart tools 2.4 plugged into Eclipse Luna 4.4.0 Integrated Development Environment (IDE) which altogether provide an integrated open-source modeling environment for model-driven development, and rapid prototyping and validation with domain experts \cite{yakindu}. Our medical models include executable models of simplified models of \textit{sepsis} and \textit{stroke} best practice guidelines consisting of both disease and underlying organ models, which are codified from medical knowledge, simplified, and then validated with physicians for correctness.

\begin{figure}[!t]
\centering
\includegraphics[width=\columnwidth]{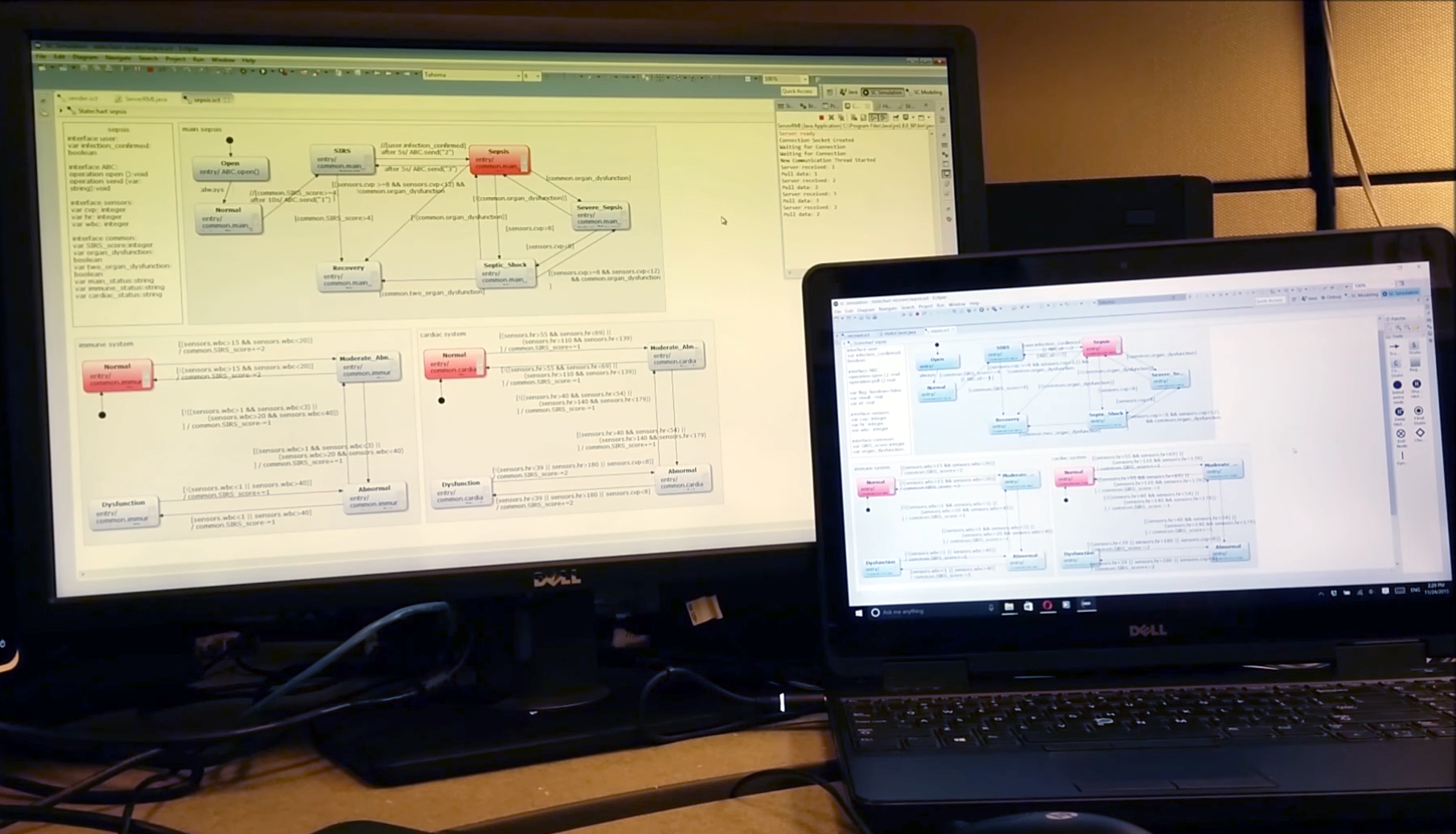}
\\
~\\
\includegraphics[width=\columnwidth]{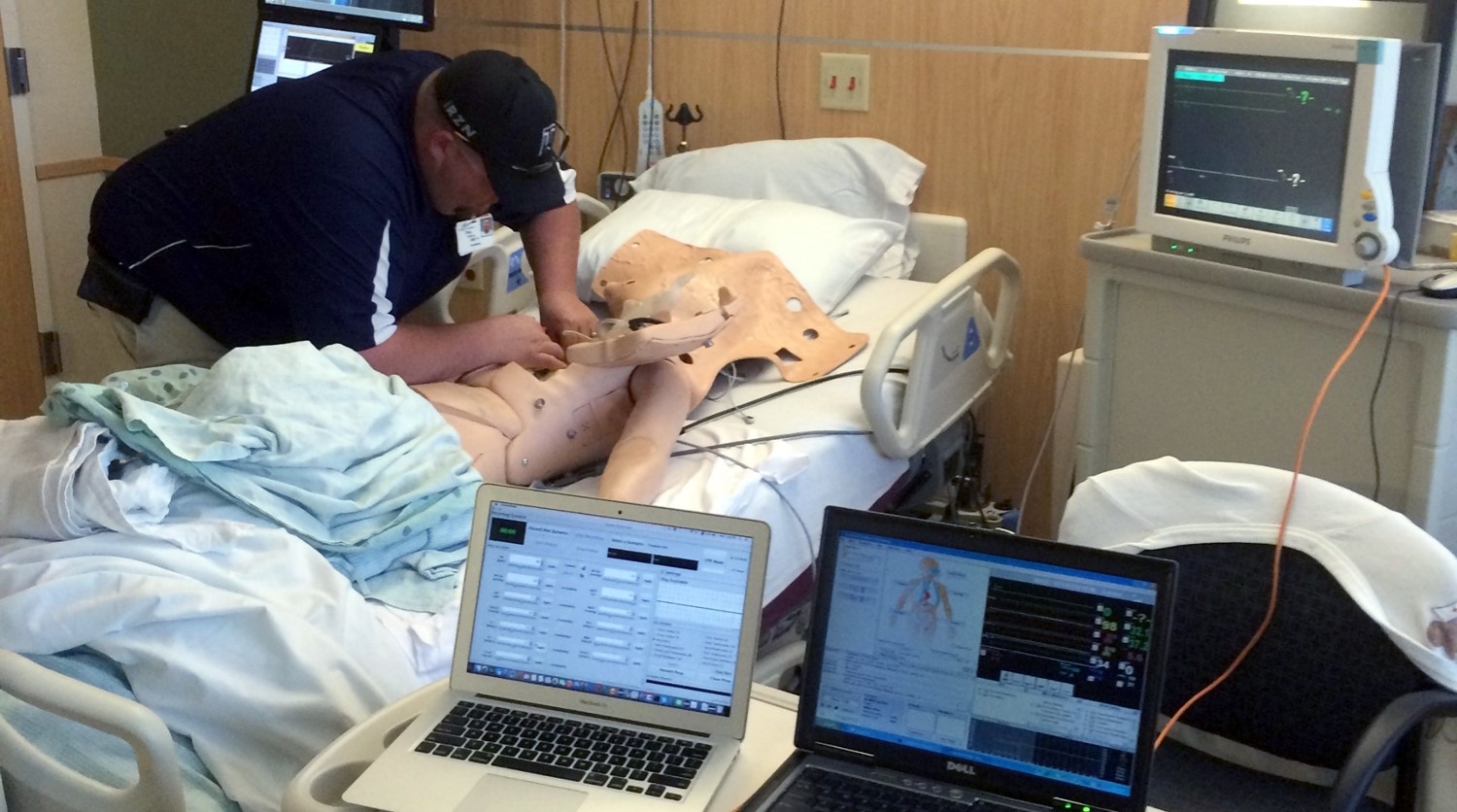}
\caption{Real-platform simulation and clinical experiments.}
\label{Fig:experiment}
\vspace{-.5cm}
\end{figure}

We have evaluated and tested the functionality of ModelSink through our proof-of-concept medical case studies conducted in collaboration with Carle Foundation Hospital~\cite{carle}. Our studies were conducted on real platforms where overall 230 types of communication and synchronization requirements were specified to provide communication across multiple sets of distributed executable medical models, and to synchronize them as necessary. Figure \ref{Fig:sepsis} illustrates an instance of simplified best-practice guidance models for sepsis, consisting of the main disease workflow model and multiple organ models. These medical statechart models are all represented as executable statecharts that focus on adherence to best-practice medical guidelines, with sets of disjoint models mounted and executed on two physically distributed machines. The machines included a Dell Latitude E5540 with Intel(R) Core i7 4600U 2.10 GHz quad-core processors, 4 MByte Cache and 8,192 MByte physical memory, running Windows 10.0 Pro 64-bit Operating System, and a HP Z230 SFF with Intel Xeon E3-1240 with 3.4 GHz quad-core processors, 8 MByte Cache and 8,192 MByte physical memory, running Windows 7 Enterprise 64-bit Operating System. Overall, the correctness of communication and synchronization operations were inspected multiple times with multi-disciplinary domain experts (10 developers, 12 researchers, and 4 physicians) to ensure that specific functional and medical requirements were satisfied and accomplished correctly. Figure \ref{Fig:experiment} presents our real-platform simulations and clinical experiments\footnote{A short simulation demo is available at:\\ http://publish.illinois.edu/mdpnp-architecture/672-2}.

\begin{figure*}[!t]
    \centering
        \includegraphics[trim=.75in 4.3in .75in 4.3in, width=.9\textwidth]{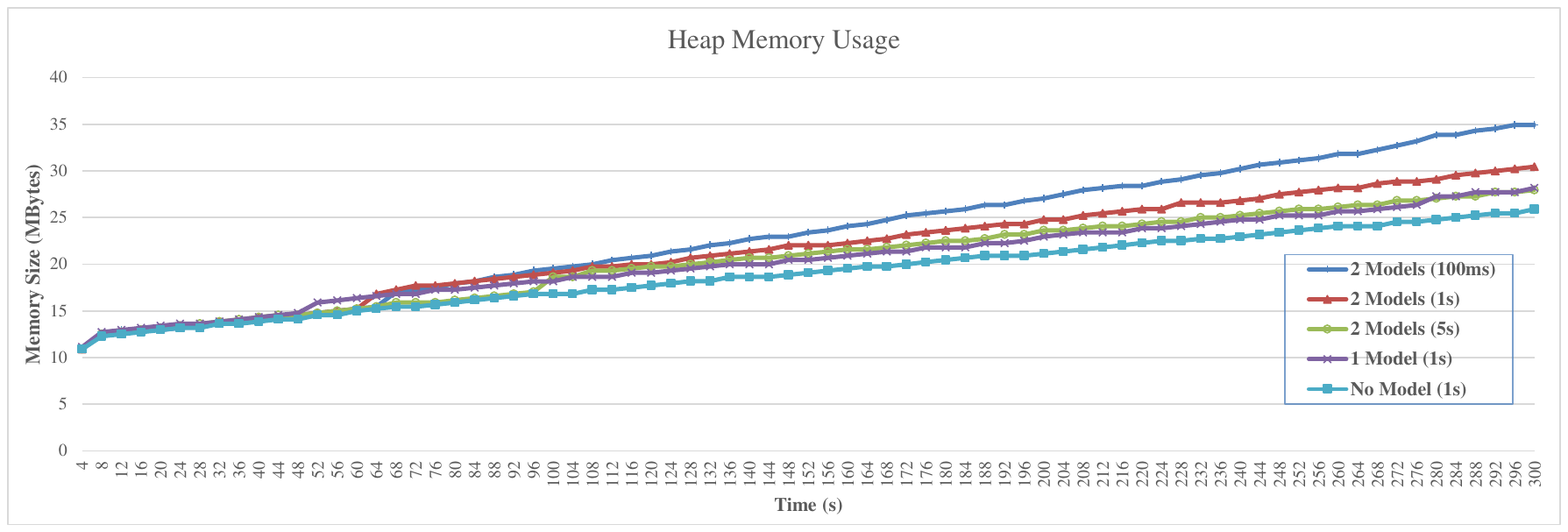}
        \includegraphics[trim=.75in 4.3in .75in 4.3in, width=.9\textwidth]{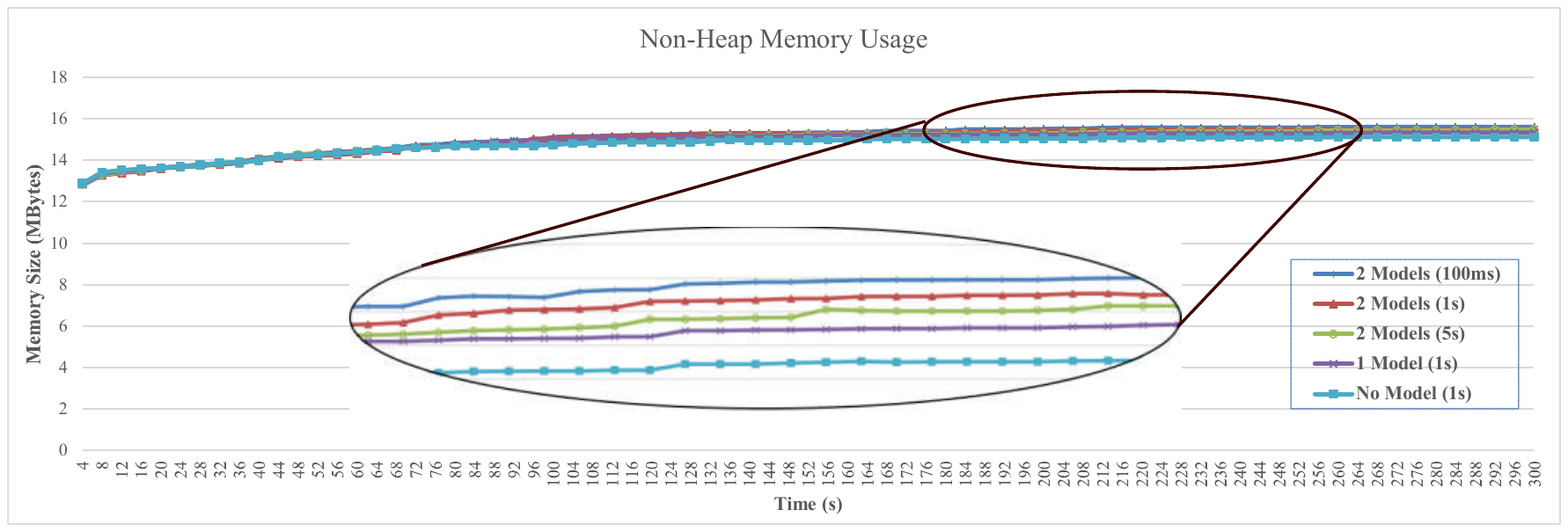}
    \caption{ModelSink's heap memory usage.}
    \label{Fig:memory}
    \vspace{-.5cm}
\end{figure*}
\subsection{Performance Evaluation}
We evaluated the performance of ModelSink through resource consumption instrumentation. We used three different monitoring tools as below for instrumentation of the Java Virtual Machine (Java VM) for fine-grained experimental data and to ensure the results are accurate and are consistent with each other:

\begin{itemize}
\item JConsole 1.8.0 \cite{jconsole}: A monitoring tool compliant with Java Management Extensions (JMX) which monitors Java Virtual Machine (JVM) and Java applications, supporting both local and remote machines.
\item VisualVM 1.38 \cite{visualvm}: A profiling tool to profile the performance and memory consumption, which provides detailed information about CPU and memory usage of Java applications.
\item JVM Monitor 3.8.1 \cite{jvmmonitor}: A Java profiling tool integrated with Eclipse as a plug-in to monitor resource usage of Java applications running on Eclipse IDE.
\end{itemize}

\begin{figure*}[!t]
    \centering
        \includegraphics[trim=.75in 4.3in .75in 4.3in, width=.9\textwidth]{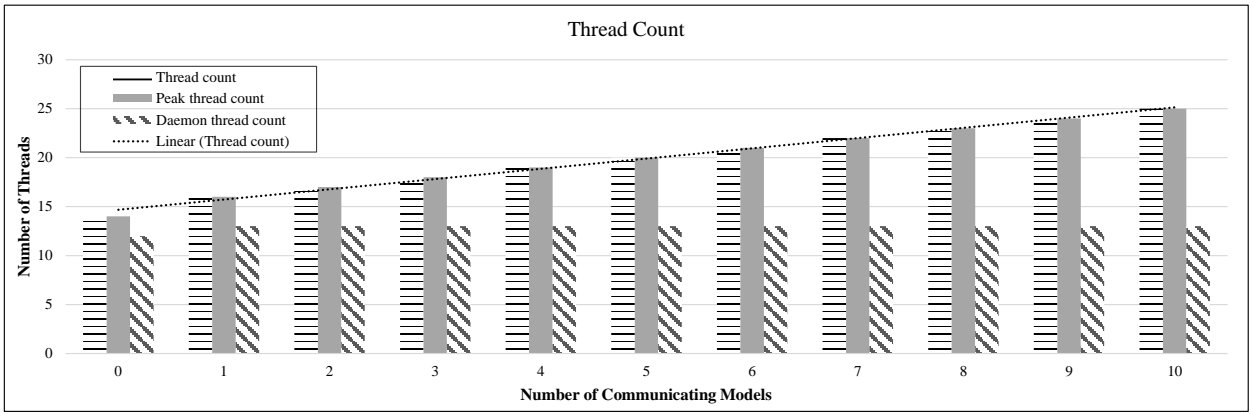}
    \caption{ModelSink's thread statistics.}
    \label{Fig:threads}
    \vspace{-.5cm}
\end{figure*}

\begin{figure*}[htbp]
    \centering
        \includegraphics[trim=.75in 4.3in .75in 4.3in, width=.9\textwidth]{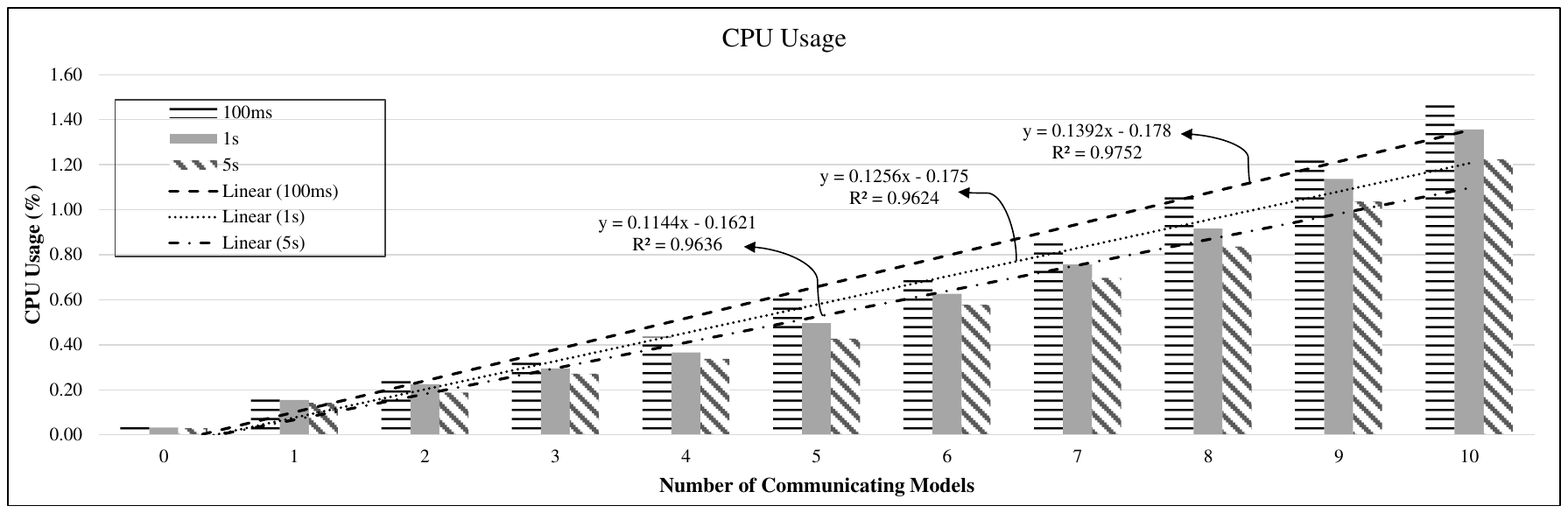}
    \caption{ModelSink's CPU usage.}
    \label{Fig:cpu}
    \vspace{-.5cm}
\end{figure*}

We employed all three profiling tools and instrumented detailed information on the CPU usage, number of threads, and heap memory consumption of ModelSink. Each profiling experiment lasted for 300 seconds, and we repeated each single experiments for 10 times to make sure the standard deviation always falls below 10\% and that the results are accurate and dependable. We tuned the polling frequency of each statechart clients at multiple rates including 100ms, 1s, and 5s to analyze trade-offs between callback frequencies and processing overhead. Figures \ref{Fig:memory} to \ref{Fig:cpu} illustrate a subset of all our performance instrumentation results.\\

\subsubsection{Memory Instrumentation}~\\
We measured the heap and non-heap memory usage of ModelSink through instrumentation done with JConsole profiler. The heap memory of JVM is created at the JVM start-up, and is the runtime data area from which memory for all class instances and arrays are allocated. We set the maximum heap size to 35MByte. The non-heap memory of JVM is also created at the JVM start-up, however, stores per-class structures and includes call stacks, memory allocated by native code for instance for off-heap caching, the Metaspace as well as memory used by the JIT compiler (compiled native code). 

Figure \ref{Fig:memory} illustrates the average heap memory consumption of ModelSink. We analyzed and compared the memory consumption of ModelSink for five different scenarios: a) baseline performance where no statechart model is communicating with ModelSink, b) a single statechart model is communicating with ModelSink, polling data with the rate of 1s and sending data as necessary, c) two statechart models are communicating with each other through ModelSink, polling data with the rate of 1s and sending data as necessary, d) two statechart models are communicating with each other through ModelSink, polling data with the rate of 100ms and sending data as necessary, and e) two statechart models are communicating with each other through ModelSink, polling data with the rate of 5s and sending data as necessary. As can be seen, the heap memory usage show an overall slight increasing linear trend due to the increase in buffer size, and no anomalies is noticed within the heap memory usage. Also, a relative comparison of the overall heap memory consumption of cases a to e above can be derived according to the figure. As can be seen, case \texttt{c} (average of 21.678 MByte over the time range of demo runtime) shows a slightly higher slope in heap memory usage compared to case \texttt{b} (average of 20.552 MByte), and case \texttt{b} shows a slightly higher slope in heap memory usage than case \texttt{a} (average of 19.234 MByte) simply due to the increase in the number of statechart clients communicating with each other. Similarly, case \texttt{e} (average of 20.579 MByte) shows a slightly lower slope in heap memory usage compared to case \texttt{c} (average of 21.678 MByte), and case \texttt{c} shows a slightly lower slope in heap memory usage than case \texttt{d} (average of 23.335 MByte) simply due to the higher rates of polling requests. Overall, an increase in the number of communicating statechart models and higher frequencies of polling requests pushes negligible overhead.

No \texttt{OutOfMemoryError} due to data spike or memory leakage is detected by the profiler during the time range of the demo runtime, approving the heap memory usage functionality of the tool. As the heap memory size reaches the defined maximum size (i.e. 35MByte), the dead objects of the heap memory are reclaimed by the JVM's garbage collection, freeing almost 31 MByte (i.e. 88\% of used memory). It is also concluded that the overhead of garbage collector process is negligible, and that garbage collector is not even called for a single point in time during a 5 minute runtime of case \textit{d} above where two statechart models are communicating through ModelSink with a polling rate of 100ms. Furthermore, in case garbage collector process runs, a single call only takes an average of 2 ms to free up heap memory, which is significantly below the polling threshold and charges relatively no performance overhead on the execution of ModelSink.

Figure \ref{Fig:memory} (middle) shows the average non-heap memory consumption of ModelSink. 
%We also instrumented the average non-heap memory consumption of ModelSink.
For non-heap memory, the abnormal growth of memory size may indicate a critical problem such as leaked loader issues or massive interned strings. In our experiments, no abnormal increase was identified in non-heap memory consumption of ModelSink over the time range of its runtime. That proves there is no class loading and object allocation problems within ModelSink. As can be seen, all figures show almost identical trendlines, with no noticeable difference. %Figure \ref{Fig:memory} (bottom) shows the same figure zoomed in significantly. 
As can be seen, a tiny difference exists among the five cases. 

Interestingly, the same argument made previously for heap memory usage is compliant with non-heap memory usage. Case \texttt{c} with 2 communicating models (average of 14.985 MByte over the time range of demo runtime) shows a slightly higher values in non-heap memory usage compared to case \texttt{b} with a single communicating model (average of 14.901 MByte), and case \texttt{b} shows a slightly higher values in non-heap memory usage than case \texttt{a} with no communicated model (average of 14.750 MByte) simply due to the increase in the number of threads corresponding to statechart clients communicating with each other and the memory stacks allocated for socket threads\footnote{Inside JVM, each Thread consumes a small amount of memory called \textit{Stack} where all active execution frames and traces are stored}. Similarly, case \texttt{e} with polling rate of 5s (average of 14.914 MByte) shows a slightly lower values in non-heap memory usage compared to case \texttt{c} with polling rate of 1s (average of 14.985 MByte), and case \texttt{c} shows a slightly lower values in non-heap memory usage than case \texttt{d} with polling rate of 100ms (average of 15.051 MByte) simply due to the higher polling rates. Overall, similar to the argument made for heap memory, it can be concluded that an increase in the number of communicating statechart models and higher frequencies of polling requests enforces negligible overhead.\\

\subsubsection{Threads Statistics}~\\
We instrumented the JVM, monitored the thread properties of ModelSink, and collected thread statistics of JVM such as the list of threads, their states, resource usage, as well as stack traces. Given the architecture design of ModelSink, a new socket thread is created and executed to serve each communicating medical thread whenever a new statechart client communicates with ModelSink. Figure \ref{Fig:threads} shows the trend in the total thread count per the total number of communicating statechart clients. We instrumented three types of JVM's monitoring statistics: a) \texttt{threadcount}, which shows the current number of live and active daemon and non-daemon threads (currently running), b) \texttt{peakthreadcount} showing the peak live thread count since the JVM started or it was reset, and c) \texttt{daemonthreadcount} showing the current number of live daemon threads. As can be seen, after the baseline with the limited number of 14 live threads, the total thread count follows a simple linear trendline, with each additional communicating statechart client initiating a new thread. No abnormal thread behavior and state change has been detected.

To identify if any deadlock is caused during the runtime of ModelSink, we inspected the states of all threads against deadlocks. Using JVM Monitor profiler \cite{jvmmonitor}, deadlocks are automatically detected and it can be recorded which threads are involved in deadlock. Our instrumentation proved no deadlock has been ever detected, and CPU was never seen to get unexpectedly overloaded.\\

\subsubsection{CPU Usage}~\\
We profiled the performance of ModelSink in terms of CPU consumption, and instrumented the CPU usage of all the threads. This is useful especially for identifying threads that have high CPU consumption. During the profiling sessions, no abnormal CPU usage was detected by a specific thread. Figure \ref{Fig:cpu} shows average CPU usage for various number of communicating statechart clients (baseline case with no communicating statechart client, up to 10 concurrent communicating statechart clients), for three different polling rates (100ms, 1s, and 5s). As can be seen, overall, the overhead of ModelSink is negligible, and no sudden spike can be noticed in the load. The average CPU consumption for 10 communicating statechart clients with polling rates of 1s are less than 5\%. The low CPU utilization of ModelSink also signifies that no source code problems such as infinite loops or excessive backend calls, and no excessive garbage collection cycles take place inside the runtime execution of ModelSink. The limited number of active threads in ModelSink helps with lowering the CPU consumption and the overall performance as the number of context switches are also limited. Figure \ref{Fig:cpu} also illustrates that the CPU consumption almost follows a linear-like trendline with high confidence (\texttt{R-Squared} value of more than 95\% fitting the linear regression lines), therefore making ModelSink scalable in terms of number of communicating statechart clients. Interestingly, the small difference in the slope of linear regression trendlines indicates that ModelSink's performance overhead is not significantly influenced by the rate of polling requests by the statechart clients, therefore making ModeSink more robust in higher polling rates.

\section{Discussion}
From a computing and software engineering perspective, within ModelSink we propose the notion of ``model-driven communication'', enabling communication among distributed models and executable state machines. Our middleware can be deployed as an add-on layer on top of modeling and decision logic software such as Yakindu's Statecharts \cite{yakindu} and Matlab's Stateflow \cite{stateflow}. ModelSink therefore assists with rapid prototyping of heterogeneous models of cyber-physical systems which are physically distributed, and makes it possible for the interaction and concurrent subjective functions such as co-design, co-simulation, and co-validation of system prototypes in various domains. Our middleware is a general purpose tool, and as long as the semantics of statechart design such as event-driven transitions is preserved, does not limit to a specific statechart tool and can be extended to other statechart and state machine design tools such as Matlab's Stateflow \cite{stateflow}, SCADE State Machine Tools \cite{scade}, and others.

In addition to the important benefits that result from the use of ModelSink, our middleware was highly praised for the automation role that it was able to play, especially in the medical domain. Prior to applying our tool, a medical technician or EMT had to coordinate manually from an ambulance transfering a patient to a regional hospital, tracing and reporting the changes, and therefore perform communication and synchronization functions manually. With the application of ModelSink, automatic communication and synchronization is achieved, removing the manual intervention of EMT, nurses, and doctors from the loop. 

Overall, we have received positive feedback from both the model-driven development industry as well as the medical staff and physicians as the end users using our middleware (the results of user study and quality of experience are prepared in a separate document). The qualitative feedback we received is promising and suggests that the middleware can in fact be applicable to large sets of requirements and that it can be extended to domains that than medical services. Such domains include large-scale co-simulation of heterogeneous production and ERP software models especially in the automotive industry~\cite{automotive}.

\section{Conclusion and Future Work}
The rapid growth of model-driven development together with distributed computer systems has led to the development of large-scale distributed best-practice statechart models. In the medical context,  the executable statechart models can assist the medical staff with clinical validation and adherence to best practices, which are achieved across a medical network from rural, through ambulance transfer, to tertiary center. However, the distributed statechart models require continuous and real-time interaction among models of different forms. This makes it necessary to offer methods of communication, especially to synchronize their execution, necessary in a medical environment.

In this paper, we describe ModelSink, a middleware that enables model-to-model communication and synchronization between heterogeneous distributed statechart models. We evaluated ModelSink, and instrumented its resource usage using medical statechart models that we have developed to assess how ModelSink performs in various loads given different performance metrics. We tested ModelSink on a real platform running distributed medical models, and demonstrated that there are in fact many potential uses of our tool in industry services other than medicine that cannot be realized through other means.

In the future, we plan to formally verify and systematically evaluate ModelSink using quantitative metrics, especially using model checking, reliability analysis, and formal verification techniques such as those proposed in \cite{jiang0}, \cite{jiang1},\cite{jiang2}, and \cite{jiang3}. We also plan to validate clinically our middleware in collaboration with Carle Foundation Hospital \cite{carle}, and to implement ModelSink on a real clinical testbed that we have built using the SimMan medical patient simulator \cite{simman}.

\section*{Acknowledgment}
This paper reviews a technology package that is the result of a team effort. Lui Sha led all system integration. Richard Berlin guided the study of patient emergency care and medical correctness. Axel Terfloth helped with the YAKINDU Statechart Tools and the semantics of model-based software development. Houbing Song helped with the safety analysis of communication protocol. Mohammad Hosseini led the development of the middleware, distribution management, communication architecture, protocol, as well as conducting the experiments and evaluations.

This research is supported in part by NSF CNS 1329886, by NSF CNS 1545002, and by ONR N00014-14-1-0717. Any opinions, findings, and conclusions or recommendations expressed in this publication are those of the authors and do not necessarily reflect the views of the grant providers.

\bibliographystyle{IEEEtran}
\bibliography{sigproc}

\begin{thebibliography}{10}
\providecommand{\url}[1]{#1}
\csname url@samestyle\endcsname
\providecommand{\newblock}{\relax}
\providecommand{\bibinfo}[2]{#2}
\providecommand{\BIBentrySTDinterwordspacing}{\spaceskip=0pt\relax}
\providecommand{\BIBentryALTinterwordstretchfactor}{4}
\providecommand{\BIBentryALTinterwordspacing}{\spaceskip=\fontdimen2\font plus
\BIBentryALTinterwordstretchfactor\fontdimen3\font minus
  \fontdimen4\font\relax}
\providecommand{\BIBforeignlanguage}[2]{{%
\expandafter\ifx\csname l@#1\endcsname\relax
\typeout{** WARNING: IEEEtran.bst: No hyphenation pattern has been}%
\typeout{** loaded for the language `#1'. Using the pattern for}%
\typeout{** the default language instead.}%
\else
\language=\csname l@#1\endcsname
\fi
#2}}
\providecommand{\BIBdecl}{\relax}
\BIBdecl

\bibitem{guideline1}
\BIBentryALTinterwordspacing
E.~J. Thomas, D.~M. Studdert, J.~P. Newhouse, B.~I.~W. Zbar, K.~M. Howard,
  E.~J. Williams, and T.~A. Brennan, ``Costs of medical injuries in utah and
  colorado,'' \emph{Inquiry}, vol.~36, no.~3, pp. 255 -- 264, Fall 1999.
  [Online]. Available: \url{http://www.jstor.org/stable/29772835}
\BIBentrySTDinterwordspacing

\bibitem{guideline2}
E.~Weise, ``Medical errors still claiming many lives,'' USA TODAY, May 2005,
  last accessed on Apr 9, 2016.

\bibitem{arden}
\BIBentryALTinterwordspacing
T.~A. Pryor and G.~Hripcsak, ``The arden syntax for medical logic modules,''
  \emph{International journal of clinical monitoring and computing}, vol.~10,
  no.~4, pp. 215--224, Nov 1993. [Online]. Available:
  \url{http://dx.doi.org/10.1007/BF01133012}
\BIBentrySTDinterwordspacing

\bibitem{glif}
V.~L. Patel, V.~G. Allen, J.~F. Arocha, and E.~H. Shortliffe, ``Representing
  clinical guidelines in glif,'' \emph{Journal of the American Medical
  Informatics Association}, vol.~5, no.~5, pp. 467--483, 1998.

\bibitem{proforma}
\BIBentryALTinterwordspacing
J.~Fox, N.~Johns, and A.~Rahmanzadeh, ``Disseminating medical knowledge: the
  \{PROforma\} approach,'' \emph{Artificial Intelligence in Medicine}, vol.~14,
  no. 1–2, pp. 157 -- 182, 1998, selected Papers from \{AIME\} '97. [Online].
  Available:
  \url{http://www.sciencedirect.com/science/article/pii/S0933365798000219}
\BIBentrySTDinterwordspacing

\bibitem{spock}
\BIBentryALTinterwordspacing
O.~Young and Y.~Shahar, \emph{Artificial Intelligence in Medicine: 10th
  Conference on Artificial Intelligence in Medicine, AIME 2005, Aberdeen, UK,
  July 23-27, 2005. Proceedings}.\hskip 1em plus 0.5em minus 0.4em\relax
  Berlin, Heidelberg: Springer Berlin Heidelberg, 2005, ch. The Spock System:
  Developing a Runtime Application Engine for Hybrid-Asbru Guidelines, pp.
  166--170. [Online]. Available: \url{http://dx.doi.org/10.1007/11527770_25}
\BIBentrySTDinterwordspacing

\bibitem{complexSoftware}
\BIBentryALTinterwordspacing
A.~Trendowicz, ``\BIBforeignlanguage{English}{Why software effort
  estimation?}'' in \emph{\BIBforeignlanguage{English}{Software Cost
  Estimation, Benchmarking, and Risk Assessment}}, ser. The Fraunhofer IESE
  Series on Software and Systems Engineering.\hskip 1em plus 0.5em minus
  0.4em\relax Springer Berlin Heidelberg, 2013, pp. 3--7. [Online]. Available:
  \url{http://dx.doi.org/10.1007/978-3-642-30764-5_1}
\BIBentrySTDinterwordspacing

\bibitem{visual}
\BIBentryALTinterwordspacing
D.~Harel, ``Statecharts: a visual formalism for complex systems,''
  \emph{Science of Computer Programming}, vol.~8, no.~3, pp. 231 -- 274, 1987.
  [Online]. Available:
  \url{http://www.sciencedirect.com/science/article/pii/0167642387900359}
\BIBentrySTDinterwordspacing

\bibitem{statecharts2}
\BIBentryALTinterwordspacing
R.~S. Moura and L.~A. Guedes, ``Using basic statechart to program industrial
  controllers,'' \emph{Computer Standards \& Interfaces}, vol.~34, no.~1, pp.
  60 -- 67, 2012. [Online]. Available:
  \url{http://www.sciencedirect.com/science/article/pii/S0920548911000687}
\BIBentrySTDinterwordspacing

\bibitem{fda1}
\BIBentryALTinterwordspacing
D.~Hoadley, ``Model-based design of medical devices,'' FDA's public workshop:
  Physiological Closed-Loop Controlled (PCLC) Devices, October 2015, last
  accessed on Apr 9, 2016. [Online]. Available:
  \url{http://www.fda.gov/MedicalDevices/NewsEvents/WorkshopsConferences/ucm457581.htm}
\BIBentrySTDinterwordspacing

\bibitem{fda2}
\BIBentryALTinterwordspacing
P.~Jones, ``Software systems assured verification,'' FDA, Last updated on March
  14, 2016, last accessed on Apr 9, 2016. [Online]. Available:
  \url{http://www.fda.gov/MedicalDevices/ScienceandResearch/ResearchPrograms/ucm477412.htm}
\BIBentrySTDinterwordspacing

\bibitem{joms}
\BIBentryALTinterwordspacing
M.~Hosseini, Y.~Jiang, P.~Wu, R.~B. Berlin, S.~Ren, and L.~Sha, ``A
  pathophysiological model-driven communication for dynamic distributed medical
  best practice guidance systems,'' \emph{Journal of Medical Systems (JOMS)},
  vol.~40, no.~11, p. 227, 2016. [Online]. Available:
  \url{http://dx.doi.org/10.1007/s10916-016-0583-5}
\BIBentrySTDinterwordspacing

\bibitem{hosseinidataset}
M.~Hosseini, R.~Berlin, and L.~Sha, ``A multi-carrier mobile geo-communication
  dataset for rural ambulance transport,'' in \emph{ACM Multimedia Systems},
  ser. MMSys '17.\hskip 1em plus 0.5em minus 0.4em\relax ACM, 2017.

\bibitem{chase17}
------, ``Adaptive clinical data communication for remote monitoring in rural
  ambulance transport,'' in \emph{ACM/IEEE International Conference on
  Connected Health}, ser. CHASE '17.\hskip 1em plus 0.5em minus 0.4em\relax
  ACM, 2017.

\bibitem{hosseiniRouteScheduler}
------, ``Physiology-aware route scheduler for emergency rural ambulance
  transport,'' in \emph{IEEE International Conference on Healthcare
  Informatics, submitted}, ser. ICHI '17.\hskip 1em plus 0.5em minus
  0.4em\relax IEEE, 2017.

\bibitem{laser1}
T.~Li, F.~Tan, Q.~Wang, L.~Bu, J.~N. Cao, and X.~Liu, ``From offline toward
  real-time: A hybrid systems model checking and cps co-design approach for
  medical device plug-and-play (mdpnp),'' in \emph{Cyber-Physical Systems
  (ICCPS), 2012 IEEE/ACM Third International Conference on}, April 2012, pp.
  13--22.

\bibitem{laser2}
F.~Tan, Y.~Wang, Q.~Wang, L.~Bu, R.~Zheng, and N.~Suri, ``Guaranteeing
  proper-temporal-embedding safety rules in wireless cps: A hybrid formal
  modeling approach,'' in \emph{Dependable Systems and Networks (DSN), 2013
  43rd Annual IEEE/IFIP International Conference on}, June 2013, pp. 1--12.

\bibitem{iccpswip}
\BIBentryALTinterwordspacing
M.~Hosseini, R.~R. Berlin, and L.~Sha, ``A physiology-aware communication
  architecture for distributed emergency medical cps,'' in \emph{Proceedings of
  the 8th International Conference on Cyber-Physical Systems}, ser. ICCPS
  '17.\hskip 1em plus 0.5em minus 0.4em\relax New York, NY, USA: ACM, 2017, pp.
  83--83. [Online]. Available: \url{http://doi.acm.org/10.1145/3055004.3064841}
\BIBentrySTDinterwordspacing

\bibitem{mediator1}
\BIBentryALTinterwordspacing
V.~Issarny, A.~Bennaceur, and Y.-D. Bromberg,
  ``\BIBforeignlanguage{English}{Middleware-layer connector synthesis: Beyond
  state of the art in middleware interoperability},'' in
  \emph{\BIBforeignlanguage{English}{Formal Methods for Eternal Networked
  Software Systems}}, ser. Lecture Notes in Computer Science, M.~Bernardo and
  V.~Issarny, Eds.\hskip 1em plus 0.5em minus 0.4em\relax Springer Berlin
  Heidelberg, 2011, vol. 6659, pp. 217--255. [Online]. Available:
  \url{http://dx.doi.org/10.1007/978-3-642-21455-4_7}
\BIBentrySTDinterwordspacing

\bibitem{mediator2}
V.~Issarny, B.~Steffen, B.~Jonsson, G.~Blair, P.~Grace, M.~Kwiatkowska,
  R.~Calinescu, P.~Inverardi, M.~Tivoli, A.~Bertolino, and A.~Sabetta,
  ``Connect challenges: Towards emergent connectors for eternal networked
  systems,'' in \emph{Engineering of Complex Computer Systems, 2009 14th IEEE
  International Conference on}, June 2009, pp. 154--161.

\bibitem{mediator3}
J.~Green, P., ``Protocol conversion,'' \emph{Communications, IEEE Transactions
  on}, vol.~34, no.~3, pp. 257--268, Mar 1986.

\bibitem{esb}
D.~Georgakopoulos and M.~P. Papazoglou, ``Enterprise service bus,'' in
  \emph{Service-Oriented Computing}, 1st~ed.\hskip 1em plus 0.5em minus
  0.4em\relax MIT Press, November 2008, pp. 1--28.

\bibitem{db1}
\BIBentryALTinterwordspacing
``Double-take share,'' Vision Solution, last accessed on Apr 9, 2016. [Online].
  Available:
  \url{http://www.visionsolutions.com/products/cross-platform-cross-data-base/double-take-share}
\BIBentrySTDinterwordspacing

\bibitem{db2}
\BIBentryALTinterwordspacing
``Data replication and fast clone,'' Informatica Data Replication, last
  accessed on Apr 9, 2016. [Online]. Available:
  \url{https://www.informatica.com/products/data-integration/real-time-integration/data-replication-and-fast-clone.html}
\BIBentrySTDinterwordspacing

\bibitem{db3}
``Symmetricds,'' \url{http://www.symmetricds.org}, last accessed on Apr 9,
  2016.

\bibitem{db4}
``Dbmoto,'' HiT Software, last accessed on Apr 9, 2016.

\bibitem{anderson}
\BIBentryALTinterwordspacing
``Clinical management algorithms,'' MD Anderson Center, last accessed on Apr 9,
  2016. [Online]. Available:
  \url{http://www.mdanderson.org/education-and-research/resources-for-professionals/clinical-tools-and-resources/practice-algorithms/clinical-management-algorithms.html}
\BIBentrySTDinterwordspacing

\bibitem{buck1994ptolemy}
J.~T. Buck, S.~Ha, E.~A. Lee, and D.~G. Messerschmitt, ``Ptolemy: A framework
  for simulating and prototyping heterogeneous systems,'' 1994.

\bibitem{jiang2015design}
Y.~Jiang, H.~Zhang, Z.~Li, Y.~Deng, X.~Song, M.~Gu, and J.~Sun, ``Design and
  optimization of multiclocked embedded systems using formal techniques,''
  \emph{Industrial Electronics, IEEE Transactions on}, vol.~62, no.~2, pp.
  1270--1278, 2015.

\bibitem{radojevic2011design}
I.~Radojevic, Z.~Salcic, and P.~S. Roop, ``Design of distributed heterogeneous
  embedded systems in ddfcharts,'' \emph{Parallel and Distributed Systems, IEEE
  Transactions on}, vol.~22, no.~2, pp. 296--308, 2011.

\bibitem{fitzgerald2014co}
J.~Fitzgerald and K.~Pierce, ``Co-modelling and co-simulation in embedded
  systems design,'' in \emph{Collaborative Design for Embedded Systems}.\hskip
  1em plus 0.5em minus 0.4em\relax Springer, 2014, pp. 15--25.

\bibitem{blochwitz2011functional}
T.~Blochwitz, M.~Otter, M.~Arnold, C.~Bausch, C.~Clau{\ss}, H.~Elmqvist,
  A.~Junghanns, J.~Mauss, M.~Monteiro, T.~Neidhold \emph{et~al.}, ``The
  functional mockup interface for tool independent exchange of simulation
  models,'' in \emph{8th International Modelica Conference, Dresden}, 2011, pp.
  20--22.

\bibitem{bombino2010heterogeneous}
M.~Bombino, M.~Hause, and P.~Scandurra, ``Heterogeneous systems co-simulation:
  a model-driven approach based on sysml state machines and simulink,'' in
  \emph{First Workshop on Hands-on Platforms and tools for model-based
  engineering of Embedded Systems, HOPES}, 2010.

\bibitem{grosso}
W.~Grosso, \emph{Java RMI}, 1st~ed., 2001, pp. 468--475.

\bibitem{openloopsafe1}
\BIBentryALTinterwordspacing
C.~Kim, M.~Sun, S.~Mohan, H.~Yun, L.~Sha, and T.~F. Abdelzaher, ``A framework
  for the safe interoperability of medical devices in the presence of network
  failures,'' in \emph{Proceedings of the 1st ACM/IEEE International Conference
  on Cyber-Physical Systems}, ser. ICCPS '10.\hskip 1em plus 0.5em minus
  0.4em\relax New York, NY, USA: ACM, 2010, pp. 149--158. [Online]. Available:
  \url{http://doi.acm.org/10.1145/1795194.1795215}
\BIBentrySTDinterwordspacing

\bibitem{openloopsafe2}
\BIBentryALTinterwordspacing
C.~Kim, M.~Sun, H.~Yun, and L.~Sha, \emph{A Medical Device Safety Supervision
  over Wireless}.\hskip 1em plus 0.5em minus 0.4em\relax Basel: Springer Basel,
  2010, pp. 21--40. [Online]. Available:
  \url{http://dx.doi.org/10.1007/978-3-0348-0031-0_2}
\BIBentrySTDinterwordspacing

\bibitem{queue1}
\BIBentryALTinterwordspacing
K.~Tan, H.~Liu, J.~Zhang, Y.~Zhang, J.~Fang, and G.~M. Voelker, ``Sora:
  High-performance software radio using general-purpose multi-core
  processors,'' \emph{Commun. ACM}, vol.~54, no.~1, pp. 99--107, Jan. 2011.
  [Online]. Available: \url{http://doi.acm.org/10.1145/1866739.1866760}
\BIBentrySTDinterwordspacing

\bibitem{queue2}
\BIBentryALTinterwordspacing
M.~M. Michael and M.~L. Scott, ``Simple, fast, and practical non-blocking and
  blocking concurrent queue algorithms,'' in \emph{Proceedings of the Fifteenth
  Annual ACM Symposium on Principles of Distributed Computing}, ser. PODC
  '96.\hskip 1em plus 0.5em minus 0.4em\relax New York, NY, USA: ACM, 1996, pp.
  267--275. [Online]. Available: \url{http://doi.acm.org/10.1145/248052.248106}
\BIBentrySTDinterwordspacing

\bibitem{organ1}
U.~Geary and U.~Kennedy, ``Clinical decision-making in emergency medicine,''
  \emph{Emergencias}, vol.~22, pp. 56--60, 2010.

\bibitem{organ2}
H.~Gao, A.~McDonnell, D.~A. Harrison, T.~Moore, S.~Adam, K.~Daly, L.~Esmonde,
  D.~R. Goldhill, G.~J. Parry, A.~Rashidian, C.~P. Subbe, and S.~Harvey,
  ``Systematic review and evaluation of physiological track and trigger warning
  systems for identifying at-risk patients on the ward,'' \emph{Intensive care
  medicine}, vol.~33, no.~4, pp. 667--679, 2007.

\bibitem{yakindu}
``Yakindu statecharts,'' http://statecharts.org, last accessed on Apr 9, 2016.

\bibitem{carle}
``{Carle Foundation Hospital},'' http://www.carle.org, last accessed on Apr 9,
  2016.

\bibitem{jconsole}
\BIBentryALTinterwordspacing
``Jconsole,'' Last accessed on Apr 9, 2016. [Online]. Available:
  \url{https://docs.oracle.com/javase/8/docs/technotes/guides/management/jconsole.html}
\BIBentrySTDinterwordspacing

\bibitem{visualvm}
\BIBentryALTinterwordspacing
``Visual vm: All-in-one java troubleshooting tool,'' Last accessed on Apr 9,
  2016. [Online]. Available: \url{https://visualvm.java.net}
\BIBentrySTDinterwordspacing

\bibitem{jvmmonitor}
\BIBentryALTinterwordspacing
``Jvm monitor: Java profiler integrated with eclipse,'' Last accessed on Apr 9,
  2016. [Online]. Available: \url{http://www.jvmmonitor.org}
\BIBentrySTDinterwordspacing

\bibitem{stateflow}
``Matlab's stateflow,'' http://www.mathworks.com/products/stateflow/, last
  accessed on Apr 9, 2016.

\bibitem{scade}
``Scade suite- control software design,''
  http://www.esterel-technologies.com/products/scade-suite/, last accessed on
  Oct 25, 2017.

\bibitem{automotive}
``Must-have erp features for the automotive industry,'' Plex Systems,
  Manufacturing Business Technology, 2014,
  http://www.mbtmag.com/articles/2014/01/must-have-erp-features-automotive-industry.

\bibitem{jiang0}
Y.~Jiang \emph{et~al.}, ``Bayesian-network-based reliability analysis of plc
  systems,'' \emph{IEEE transactions on industrial electronics}, vol.~60,
  no.~11, pp. 5325--5336, 2013.

\bibitem{jiang1}
Y.~Jiang, M.~Gu, J.~Sun, and L.~Sha, ``Data-centered runtime verification of
  wireless medical cyber-physical system,'' \emph{IEEE transactions on
  industrial informatics}, 2016.

\bibitem{jiang2}
Y.~Jiang \emph{et~al.}, ``Design of mixed synchronous/asynchronous systems with
  multiple clocks,'' \emph{IEEE Transactions on Parallel and Distributed
  Systems}, vol.~26, no.~8, pp. 2220--2232, 2015.

\bibitem{jiang3}
------, ``Design and optimization of multiclocked embedded systems using formal
  techniques,'' \emph{IEEE Transactions on Industrial Electronics}, vol.~62,
  no.~2, pp. 1270--1278, 2015.

\bibitem{simman}
``Simman patient simulator, laerdal medical,''
  \url{http://www.laerdal.com/doc/86/SimMan}, last accessed on Apr 9, 2016.

\end{thebibliography}


\begin{IEEEbiography}[{\includegraphics[width=1in,height=1.25in,clip,keepaspectratio]{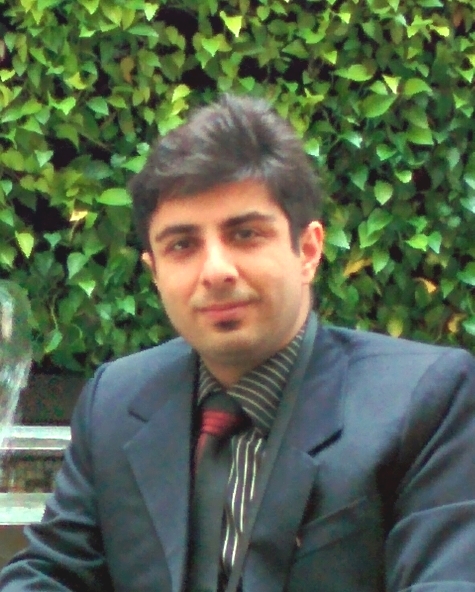}}]{Mohammad Hosseini}
received the Ph.D. degree in Computer Science from University of Illinois at Urbana-Champaign (UIUC) in 2017. His research background mainly lies in systems and networking area, including medical CPS. He has served as TPC member, reviewer, and guest reviewer at major workshops, conferences, and journals including IEEE TMM, ACM MM, ACM MMSys, IEEE ICME, ACM MoVid, IEEE Netgames, Springer MMSJ, etc. Dr. Hosseini was a winner of the prestigious 2017 IEEE Communication Society (IEEE ComSoc) Competition Award.
\end{IEEEbiography}

\begin{IEEEbiography}[{\includegraphics[width=1in,height=1.25in,clip,keepaspectratio]{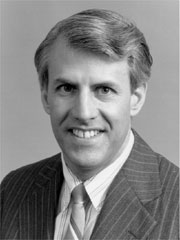}}]{Richard R. Berlin}
Richard Berlin received his MD from SUNY-Downstate in 1974 and MBA from CUNY-Baruch in 1992.​ Dr. Richard Berlin is a Level 1 trauma surgeon at Carle Foundation Hospital, and an adjunct associate professor in the Department of Computer Science at UIUC. He served as the medical director for health systems at Health Alliance, the regional Health Maintenance Organization (HMO) for central Illinois.  The co-author of the text Healthcare Informatics (Springer), Dr. Berlin's interests include Systems Medicine summarized in a most recent (Springer journal) publication, Systems Medicine - Complexity Within, Simplicity Without.
\end{IEEEbiography}

\begin{IEEEbiography}[{\includegraphics[width=1in,height=1.25in,clip,keepaspectratio]{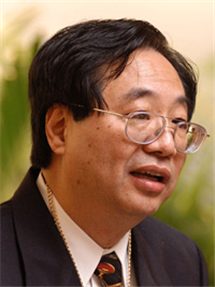}}]{Lui Sha}
received the Ph.D. degree from Carnegie Mellon University, Pittsburgh, PA, in 1985. He is currently a Donald B. Gillies Chair Professor of computer science at the University of Illinois at Urbana Champaign. His work on real-time computing is supported by most of the open standards in real-time computing and has been cited as a key element to the success of many national high-technology projects including GPS upgrade, the Mars Pathfinder, and the International Space Station. Professor Sha is a recipient of the prestigious 2016 IEEE Simon Ramo Medal for technical leadership and contributions to fundamental theory, practice, and standardization for engineering real-time systems.
\end{IEEEbiography}

\begin{IEEEbiography}[{\includegraphics[width=1in,height=1.25in,clip,keepaspectratio]{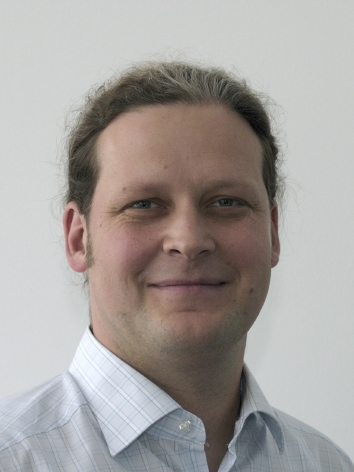}}]{Axel Terfloth}
heads research and development for embedded systems at itemis AG, Germany. He deals with the adaption and development of methods and technologies for integrated tool chains and model-based development for more than 20 years now. He is involved in different commercial and research projects, and is responsible for the modeling tool Yakindu Statechart Tools. He regularly writes and speaks on these subjects.
\end{IEEEbiography}

\begin{IEEEbiography}[{\includegraphics[width=1in,height=1.25in,clip,keepaspectratio]{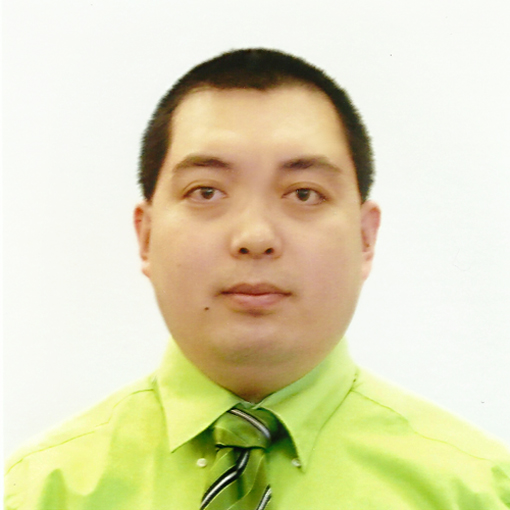}}]{Houbing Song}
received the Ph.D. degree in electrical engineering from the University of Virginia, Charlottesville, VA, in 2012. He is currently an Assistant Professor and the Director of the Security and Optimization for Networked Globe Laboratory (SONG Lab) at the Department of Electrical, Computer, Software, and Systems Engineering, Embry-Riddle Aeronautical University, Daytona Beach, FL. He serves as an Associate Technical Editor for IEEE Communications Magazine. He is the editor of multiple books, and has authored more than 100 articles. His research interests include cyber-physical systems, cybersecurity and privacy, internet of things, connected vehicle, smart and connected health, and wireless communications and networking.

Dr. Song is a senior member of ACM. Dr. Song was the very first recipient of the Golden Bear Scholar Award, the highest campus-wide recognition for research excellence at West Virginia University Institute of Technology (WVU Tech), in 2016. 
\end{IEEEbiography}

\end{document}